\documentclass[11pt,oneside,letterpaper,table]{article}
\usepackage[utf8x]{inputenc}
\usepackage[letterpaper,margin=1in]{geometry}
\usepackage{setspace}
\usepackage[T1]{fontenc}
\usepackage[sc]{mathpazo}
\usepackage{nomencl} 			
\usepackage{graphicx}			
\usepackage{subfig}
\usepackage{microtype} 		
\usepackage{multirow}
\usepackage{array}
\usepackage{tabulary}
\newcolumntype{K}[1]{>{\centering\arraybackslash}p{#1}}
\usepackage{bigstrut}
\usepackage{rotating}
\usepackage{booktabs}
\usepackage{float}
\usepackage{amsmath}
\usepackage{nccmath}
\usepackage{rotating}
\usepackage{longtable} 
\usepackage{pdflscape}
\usepackage{epstopdf}
\usepackage{threeparttablex}
\usepackage{caption}
\usepackage{dcolumn}
\usepackage{amssymb}
\usepackage{array}
\newcolumntype{C}[1]{>{\centering\arraybackslash}p{#1}}
\newcolumntype{P}[1]{>{\arraybackslash}p{#1}}
\usepackage{hhline}
\usepackage{natbib}
\usepackage[colorlinks=true]{hyperref}
\usepackage[font=normalsize,labelfont=bf]{caption}
\usepackage[bottom,multiple]{footmisc}
\renewcommand{\arraystretch}{1.3}
\usepackage{titling}
\usepackage{tikz}
\usepackage{subfig}
\usepackage{relsize}
\usepackage{lipsum}
\usepackage{adjustbox}
\usepackage{svg}
\usepackage{graphicx}
\usepackage{calc}
\newlength{\depthofsumsign}
\usepackage{setspace}
\usepackage{etoolbox}
\newbox\globalTPTbox
\usepackage{xcolor}
\usepackage{pgfplotstable}
\usepackage{pgf}
\usepackage{collcell}
\usepackage{booktabs}

\usepackage[page,toc,titletoc,title]{appendix}
\usepackage{minitoc}

\newcommand*{\MinNumber}{0}%
\newcommand*{\MaxNumber}{0.5}%

\newcommand{\ApplyGradient}[1]{%
  \pgfmathsetmacro{\PercentColor}{65*(#1-\MinNumber)/(\MaxNumber-\MinNumber)}%
  \edef\x{\noexpand\cellcolor{black!\PercentColor}}\x\textcolor{black}{#1}%
}
\newcolumntype{R}{>{\collectcell\ApplyGradient}{c}<{\endcollectcell}}



\makeatletter
\newcommand{\fixsubtable}{\let\caption=\caption@caption
  \def\@captype{table}}
\makeatother

\title{Predicting Inflation with Recurrent Neural Networks}
\author{\Large Livia Paranhos\thanks{E-mail: \texttt{Livia.Silva-Paranhos@bankofengland.co.uk}. The views expressed are solely those of the author and do not represent the views of the Bank of England.} \\ Bank of England}
\date{\small First version: November 2020 \\ This version: September 2023}

\begin{document}

\hypersetup{%
    ,citecolor=blue
    ,linkcolor=red
    }

\maketitle
\thispagestyle{empty}

\begin{abstract} \small

This paper applies a recurrent neural network, the LSTM, to forecast inflation. This is an appealing model for time series as it processes each time step sequentially and explicitly learns dynamic dependencies. The paper also explores the dimension reduction capability of the model to uncover economically-meaningful factors that can explain the inflation process. Results from an exercise with US data indicate that the estimated neural nets present competitive, but not outstanding, performance against common benchmarks (including other machine learning models). The LSTM in particular is found to perform well at long horizons and during periods of heightened macroeconomic uncertainty. Interestingly, LSTM-implied factors present high correlation with business cycle indicators, informing on the usefulness of such signals as inflation predictors. The paper also sheds light on the impact of network initialization and architecture on forecast performance.

\vspace{5mm}


\end{abstract}

\newpage

\doparttoc 
\faketableofcontents 

\section{Introduction}

A good forecast model for inflation is essential for economic agents and policy makers, yet up until recently, it was hard to improve over simple univariate specifications (\citealp{ao}; \citealp{stock07}; \citealp{faust}). New advances in machine learning have spurred the interest of economists in applying different methods to macroeconomic forecasting with the general promise of improvements in forecast accuracy. The common downside of machine learning techniques, including neural networks, is the lack of interpretability. For inflation prediction in particular this could be a problem, since much of the effort is devoted to understand the (likely unstable) underlying inflation process, sometimes at the expense of marginal increases in forecasting gains.

This paper investigates the ability of a special class of neural networks, the recurrent networks, to forecast inflation, and explores how these models can inform about the inflation process. While neural networks are well known by their flexibility in modelling complex data processes, recurrent networks were specifically designed to model sequences of observations, which makes them appealing for the study of time series. I consider specifically the LSTM (short for ``long-short term memory''), a variation of the plain recurrent network, as the model of interest. This type of model differs from fully connected feed-forward networks in the way it handles past information. While the latter process all lags forwards (as its name suggests) and simultaneously in the network, recurrent models process each time step sequentially which enables the model to learn the dynamic dependencies in the data. LSTM networks in particular are able to learn from time steps far back in time, as opposed to plain recurrent models. In practice, this is achieved by incorporating a number of filters in the network that control the flow of information across time. This so called \emph{long-memory} characteristic may be particularly important to model the long term trend of the series.

The paper also explores how the LSTM is useful as a nonlinear dimension reduction tool, which is in general achieved by carefully designing the network architecture (\citealp{Goodfellow-et-al-2016}), to uncover economically-meaningful latent states that are behind the inflation process. The outcome can roughly be interpreted as a nonlinear generalization of common factors extracted from Principal Component Analysis (PCA). The difference is that, given the predictive nature of the model, the factors are in practice learned by supervision with the target variable, and should therefore carry information on the drivers of the inflation process.

I compare the forecasting performance of LSTM-based models with the fully connected feed-forward network (with varying input sets) and a number of benchmarks, including other machine learning models (Random 
Forest, LASSO, Ridge, and Elastic-net), as well as more traditional models in the inflation forecasting literature. I consider forecasting the US Consumer Prices Index (CPI) inflation, with extended results on Personal Consumption Expenditures (PCE) inflation in the Appendix.

\noindent \textbf{RESULTS.} On forecast performance, the estimated neural networks show competitive results against traditional benchmarks and other machine learning models, but these are not outstanding, in particular when compared to the random forest. The LSTM appears a good model to forecast at long horizons, which can be explained by its plausible ability in modelling long term trends, as well as during periods of instability, as it is relatively insulated from sudden, short-lived movements in inflation. Results for feed-forward models are more mixed, as these change with the horizon and input data, although their performance remains competitive against the benchmarks. Additionally, network architecture (here referring to the number of nodes in the network) is found to influence forecast performance. Smaller networks for which the number of nodes was \emph{ex-ante} fixed show satisfactory, although generally worst, performance compared to bigger, cross-validated networks. This result encourages the use of ensemble techniques that account for multiple types of network architectures whenever the interest is mainly on forecast accuracy.

On inflation modelling, the LSTM reveals to be an interesting tool to reduce the dimensionality of the data in a way that is relevant for prediction. When estimated with the US FRED-MD database (\citealp{fred}), the estimated LSTM factors appear to capture well the underlying inflation trend and exhibit high correlation with business cycle indicators, in particular with the output gap, informing on the usefulness of such signals as inflation predictors themselves. Additionally, a variance decomposition analysis reveals that LSTM factors are significantly loaded on housing starts and corporate bond spreads, and they generally agree with the literature on common predictors of inflation.

In addition to make empirical contributions on the usefulness of these models for inflation modelling and forecasting, the paper also informs on more practical issues related to its implementation. In this respect, I discuss the impact of network initialization on forecast performance, as it is known that neural network predictions can be quite sensitive to initial conditions. According to a standard \cite{diebold}-type statistic, I show that the ensemble prediction---the average prediction of models embedding different initial values---is at least as good as the prediction of networks with a specific initialization, a result that is reassuring given the widespread use of this technique for this class of models. Additionally, the exercise also reveals that predictions from networks with different initial values are highly heterogeneous, which suggests that in practice the algorithm usually gets stuck into local optima, which again supports the use of ensembles.

\noindent \textbf{RELATED WORK.} This paper relates more broadly to the literature on inflation forecasting, and more specifically to papers applying neural networks to inflation or macroeconomic forecasting. The paper also relates to the vast literature on dimensional reduction techniques, where here I discuss the narrower literature on the use of deep neural nets for this purpose.

The conventional literature on inflation forecasting usually refers to Phillips curve-based models that relate inflation to some activity variable and autoregressive terms (\citealp{stock09} provide an overview). Although well-established, forecasts based on the (more traditional) Phillips curve have varying performances over time and can be quickly overperformed by univariate models, as the random walk (\citealp{ao}), the unobserved components model (\citealp{stock07}), or autoregressive models more generally. Other commonly used benchmarks are the Bayesian vector autoregression 
 model (\citealp{giannone}) and dynamic factor models (\citealp{stock02}; \citealp{lud}).

The recent improvements in machine learning have shifted the attention of econometricians to new, promising techniques for macroeconomic forecasting (\citealp{EXTERKATE}; \citealp{joseph}; \citealp{medeiros}; \citealp{coulombe2020machine}).\footnote{In the field of economics more broadly, see for example \cite{mullainathan} and \cite{athey} for an assessment of those methods applied to policy analysis and causal inference respectively, \cite{varian} for a discussion about big data in economics, and \cite{refenes&white} and \cite{gu} for applications in finance.} For instance, in an extensive exercise to forecast US CPI inflation, \cite{medeiros} point to large improvements in forecast accuracy of machine learning models incorporating a large pool of predictors, especially the random forest model. The forecasting exercise of the present paper is roughly similar to theirs in terms of sample, data and the benchmark models considered (although their paper is more extensive on this), the difference being that here I extend the analysis to neural networks and recurrent networks in particular. 

Regarding the specific use of neural networks for inflation and macroeconomic prediction, this paper is part of a modest yet growing literature. Early papers are e.g. \cite{kuan}, \cite{nakamura} and \cite{MCADAM}. With the advances in computing power and big data, this literature grew, with examples considering feed-forward neural networks (\citealp{choudhary}; \citealp{joseph}; \citealp{coulombePC}), plain recurrent neural networks (\citealp{sermpinis}), and LSTM networks (\citealp{cook}; \citealp{almosova}; \citealp{verstyuk}). On LSTM networks in particular, \cite{cook} compare the forecasting performance of different network models, including a deep neural network, a convolutional network, the LSTM, and an encoder-decoder network, in a univariate exercise for US unemployment. \cite{verstyuk} compares a small-scale LSTM model to vector autoregression specifications in terms of forecasting accuracy and impulse response dynamics. \cite{almosova} is the paper that most resembles the present one. The authors compare the forecasting performance of the LSTM with a feed-forward network and linear benchmarks in an exercise with US CPI. In terms of accuracy, they also point to a better performance of the LSTM at long horizons and highlight the importance of network architecture on the results. One difference is that they consider a univariate framework, while the present paper explores the use of big data and dimensional reduction. Additionally, this paper presents a wider set of benchmark models which includes other widely applied machine learning models, as the random forest and shrinkage methods.

Finally, this paper also touches the field of dimension reduction. In addition to prediction tasks, deep learning is also an interesting tool for nonlinear compression and can be regarded as a generalization of PCA (\citealp{hinton}). The use of this technique has been growing recently, with applications using Encoder-Decoder structures in an unsupervised fashion (\citealp{andreini}; \citealp{hauzenberger}), or in a supervised, forecasting framework (\citealp{coulombePC}). This work relates more closely in this regard to \cite{coulombePC}. The author uses a carefully designed deep neural net to extract \emph{ex-ante} unobserved components of the Phillips curve in a supervised exercise to forecast inflation. The present paper does not impose the linear structure inherent to the Phillips curve as in  \cite{coulombePC}, and extracts economically-interpretable factors via a careful selection of predictor sets. It is worth mentioning the relation between the LSTM and Encoder-Decoder models as dimension reduction tools. In the latter, the encoder is responsible for compressing the original data into a generally small number of components (the factors) and the decoder for relating these factors to the target (in the case of Autoencoders, the target would be the original data itself). The LSTM model applied in this paper can be seen as a variant of this type of model, where a first block containing the LSTM structure plays the role of the encoder (as it maps the original data into a low-dimensional state), and a second block involving a deep feed-forward network operates as the decoder, linking the information from the factors to the target variable. Although it is not \emph{ex-ante} clear what are the differences between the implied factors of each method, it is likely that LSTM factors give more emphasis to dynamic dependencies in the data and less to other forms of nonlinearities arising between predictors, and that Encoder-Decoder factors are better suited to capture these type of nonlinearities as per the deep network structure of the encoder.

\noindent \textbf{OUTLINE.} The organization of the paper is as follows. Section \ref{section:framework} introduces the econometric framework and the neural network models. Section \ref{sec:estimation} discusses model selection while Section \ref{sec:data} presents the data. Section \ref{sec:comp} investigates the properties of the estimated LSTM factors. Section \ref{sec:forecasting} discusses out-of-sample forecast performance, and Section \ref{sec:iv} the sensitivity analysis over network initialization. Section \ref{sec:conclusions} concludes.

\section{Econometric framework}
\label{section:framework}

Consider the set of macroeconomic predictors $\mathcal{Z} = (w_t,z_t)'$, where $w_t$ is a vector collecting the inflation series to be forecast as well as disaggregate inflation components, and $z_t$ is a vector collecting macroeconomic series excluding $w_t$. Importantly, the set $w_t$ is not contained in $z_t$, which allows me to isolate the predictive effect of the macroeconomic variables on inflation. Without loss of generality, I set the first element of $w_t$ as the inflation series to be forecast, and denote it by $y_t$.

\sloppy Let $x_t$ be the $NL$-size vector collecting the predictors at time $t$ and their lags. In this paper, I consider predictor sets of the form 
$x_t = (z_t,...,z_{t-(L-1)})'$, $x_t = (w_t,...,w_{t-(L-1)})'$ or $x_t = (z_t,...,z_{t-(L-1)},w_t,...,w_{t-(L-1)})'$, where $N$ changes accordingly across sets. I assume that the $h$-step ahead inflation $y_{t+h}$ evolves nonlinearly with respect to the predictors $x_t$ through the function $G$, such that
\begin{equation}
    y_{t+h} = G(x_t;\Theta_h) + \varepsilon_{t+h}.
\label{eq:main_spec}
\end{equation}
$\Theta_h$ collects the model parameters and $\varepsilon_{t+h}$ is the prediction error. The underlying statistical problem consists therefore in estimating the unknown function $G: x_t \mapsto y_{t+h}$.

In this application, $G$ takes the form of a neural network, in which case fitting the function to the data corresponds to estimating $\Theta_h$ given a network architecture. An architecture $\mathcal{A}_G$ is specified as being a collection of choices that defines the functional form of $G$. It embeds two elements, (i) the neural network model (or a combination of neural network models), and (ii) hyperparameters specific to the model. The network models considered in this paper are the feed-forward network, the recurrent network with LSTM units---referred here as the LSTM model---, and a combination of both, the FF-LSTM, as detailed below. Importantly, the choice of the network is defined \emph{ex-ante} by the researcher while the hyperparameters are selected via cross-validation. Making an analogy to the nonparametric literature, the hyperparameters can be viewed as tuning parameters that are model-specific, e.g. the bandwidth in kernel regression, or the choice of $k$ in $k$-nearest-neighbors estimation. 

Let $\mathcal{S}_{\mathcal{A}_G}$ be the set of parameters specific to architecture $\mathcal{A}_G$. The parameters $\Theta_h \in \mathcal{S}_{\mathcal{A}_G}$ are estimated by minimizing the mean squared error loss
\begin{equation}
   \hat{\Theta}_h = \underset{\Theta_h \in \mathcal{S}_{\mathcal{A}_G}}{arg min} \left\{ \frac{1}{T-h} \sum_{t=1}^{T-h} \Big ( y_{t+h} - G(x_t;\Theta_h) \Big )^2 \right\}
\label{eq:loss}
\end{equation} 
for a given set of predictors $x_t$ and target variable $y_{t+h}$, where the estimation is implemented by gradient descent (more details in Appendix \ref{app:estimation}).

The next sections present a detailed description of the three models considered. They mainly differ from one another with respect to the underlying network structure and the predictor set.

\subsection{The feed-forward model}

\begin{figure}
    \centering
    \caption{The feed-forward neural network}
    \includegraphics[width=9cm]{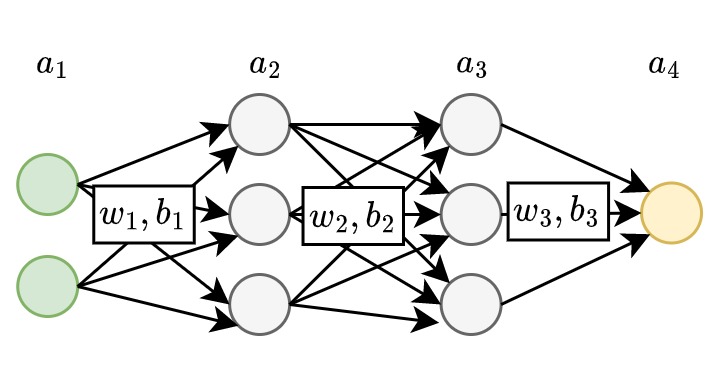}
    \caption*{\small Schematic representation of a feed-forward neural network with $q+1=4$ layers, $n=3$ nodes per hidden layer, $2$ inputs and $1$ output.}
    \label{fig:ff}
\end{figure}

This is the plain fully connected feed-forward (FF) network. It consists of a potentially large number of simple elements (nodes) that are organized into layers: (i) a first layer that receives the raw input information, (ii) a final layer that corresponds to the network predictions, and (iii) hidden layers that lie in between, processing information in one direction, from the first to the last layer. Figure \ref{fig:ff} provides a schematic representation. Each node in a given layer receives a linear combination of the outputs $a$ from the previous layer in the network,
$a'w+b$, where $w$ and $b$ are parameters, and returns a nonlinear transformation of it, $\sigma(a'w+b)$. The estimation of the model consists of learning the parameters in the network, also known as weights, using gradient descent (details in Appendix \ref{app:estimation}).

I consider using the rectified linear unit (ReLu), $z \mapsto max\{0,z\}$, as the nonlinear transformation applied at each node, following modern practices in deep learning. Recent works on deep neural networks (networks with a large number hidden layers) establish formal results on the statistical properties of ReLu-based networks and derive faster rates of convergence compared to networks using sigmoid-based functions (\citealp{schmidt}; \citealp{farrell}). These works follow the enormous success of ReLu-based deep nets in prediction tasks which are now becoming the state of the art in many contexts.

More formally, consider a feed-forward neural network with $q+1$ layers in total. Let $a_i$ be the hidden layer containing $n$ nodes, for $i=2,...,q$.\footnote{The restriction that each layer must contain the same $n$ number of nodes is imposed for simplification.} The feed-forward model takes the form
\begin{equation}
    \begin{aligned}
        & g_{FF}(x_t) = w_{q}'a_q + b_{q} \\
        & a_i = ReLu(w_{i-1}' a_{i-1} + b_{i-1}), \quad i=2,...,q \\
        & a_1 = x_t,
    \end{aligned}
\label{eq:ff}
\end{equation}
where $\{w_i,b_i\}_{i=1}^q$ are the parameters of the model. Hence for the FF model, $G(x_t;\Theta_h) = g_{FF}(x_t)$. The dimension of each parameter depends on the layers it connects. Specifically, $w_1 \in \mathbb{R}^{NL\times n}$, $\{w_i\}_{i=2}^{q-1} \in \mathbb{R}^{n \times n}$, and $w_q \in \mathbb{R}^{n\times 1}$, while for the intercept terms, $\{b_i\}_{i=1}^{q-1} \in \mathbb{R}^{n\times 1}$ and $b_q$ is a scalar. The hyperparameters of this model are the number of layers $q$, the number of nodes per layer $n$ and the number of lags $L$ to include in the input vector $x_t$. Note here that the size of the output layer is fixed to one given the single-valued target variable.

I consider estimating this model with both inflation-only data, $x_t = (w_t,...,w_{t-(L-1)})'$, as well as with the pool of economic predictors, $x_t = (z_t,...,z_{t-(L-1)})'$. The first case, called FF-cpi, is a natural nonlinear extension of the autoregressive model that maps lags of inflation (and its components) to the $h$-step ahead inflation. The second case, called FF-pool, forecasts inflation with a large set of macroeconomic predictors (and its lags). The total number of parameters of the feed-forward model is $(NL+1)n + (Q-1)(n+1)n + (n+1)$.

\subsection{The LSTM model}
\label{section:macro_model}

The LSTM is a variant of the recurrent neural network. This type of model differs from FF networks in the way it handles past information. While FF models process all the input lags forwards and simultaneously in the network, recurrent networks process each time step sequentially, allowing the output of a previous time step to be the input of the following one. Note that in this case the information flows across the lag structure of the data, while in FF models it essentially flows across layers where the dynamic structure is not explored explicitly. This feature makes recurrent neural nets quite attractive to model time series, as they ``remember'' the information contained in previous time steps and may use it to improve prediction accuracy.

The main insight from the recurrent structure is a so called \emph{internal memory} that is updated at each time step. The model then estimates the parameters such that its internal memory embeds the relevant information to forecast the target, but importantly it does so by taking into account the sequential dependency inherent in the data. For example, if low inflation is associated with a low output gap that persists during a period of time, but not with a low output gap that is essentially transitory, then the model is in principle able to learn these dynamic dependencies and extrapolates it to the forecast. An additional interesting feature of these models is their ability to generate factor-like estimates that are relevant for predicting inflation. This property is alike as the supervised nature of partial least squares as an alternative to principal component regression, which is by definition unsupervised. This essentially follows from the fact that the parameters of the LSTM model are learned by supervision with inflation, and consequently the components of the internal memory as well. In the rest of the paper, I interpret and refer to the internal memory as \emph{LSTM factors} in an analogy to factor analysis.

\begin{figure}
    \centering
    \caption{The LSTM and the flow of information}
    \includegraphics[width=12cm]{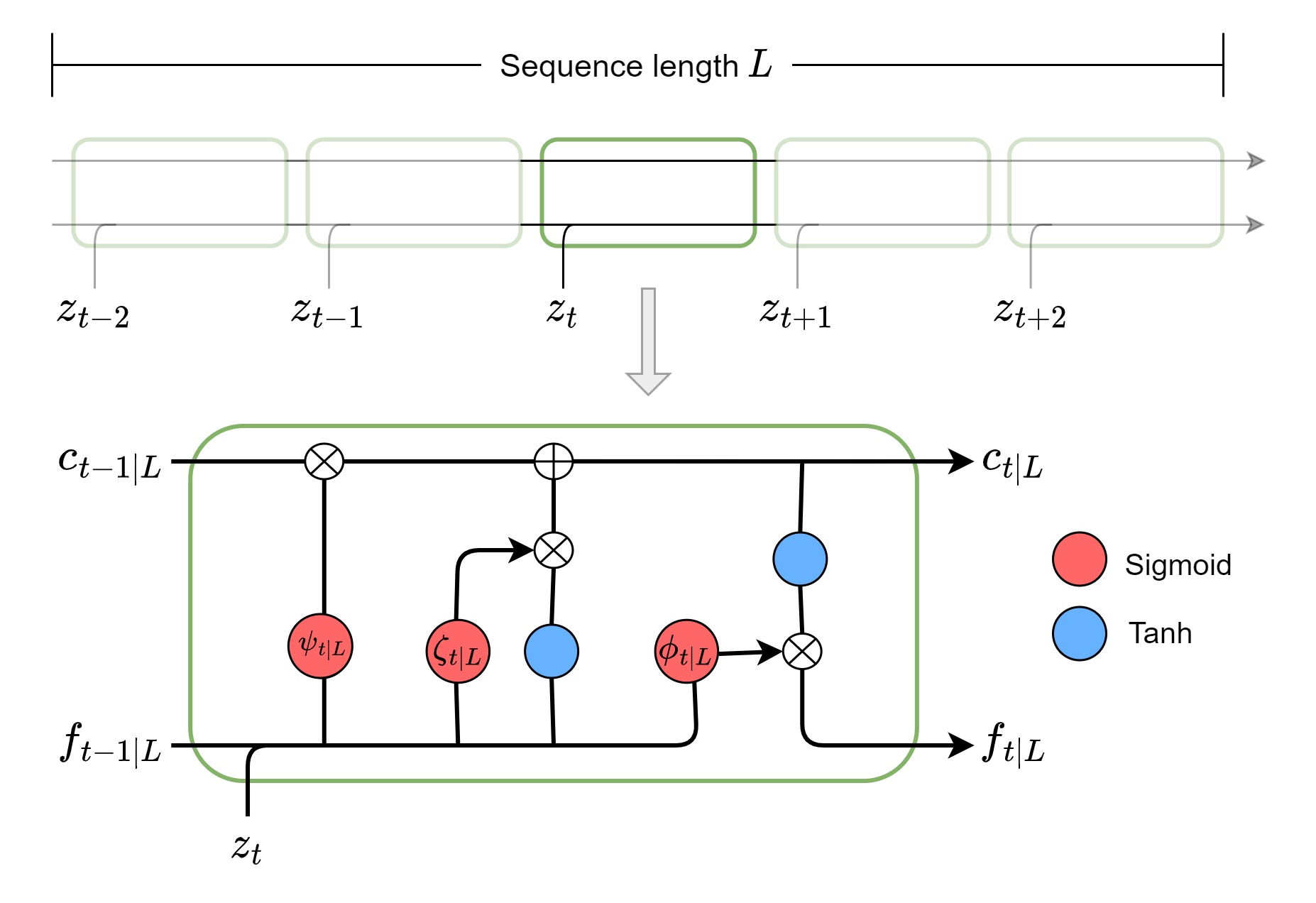}
    \caption*{\small The figure is a visual representation of the LSTM equations in (\ref{eq:lstm}). Symbols $\oplus$ and $\otimes$ denote element-wise addition and multiplication respectively.}
    \label{fig:lstm_diag}
\end{figure}

To be more concrete, consider e.g. the predictor set $x_t = (z_t,...,z_{t-(L-1)})'$ collecting the $N\times 1$ vector of predictors and their lags. The LSTM factors are represented as a $p$-vector denoted $f_t$ and are a function of both the current input information $z_t$ and the lagged factors $f_{t-1}$.
Importantly, this recursion is limited to a fixed lag $L$ in a way that past information can only be traced back up to $L$. In order to embed this idea in the notation, I write $f_{t|L} \equiv f_{t|t,t-1,...,t-(L-1)}$. In practice, the recursion behind the LSTM takes into account an additional component (the ``cell state'' in machine learning jargon) that ensures that the algorithm is able to learn from time steps far away in the past. This is in contrast to a plain recurrent neural net that suffers form vanishing gradients (more details on plain recurrent nets are in Appendix \ref{app:RNN}). The ultimate structure of the LSTM then combines the cell state, denoted $c_{t|L}$, with the LSTM factors $f_{t|L}$ through a number of filters ($\psi_{t|L}, \phi_{t|L}, \zeta_{t|L}$) that essentially control the flow of information through time. These filters are defined as sigmoid functions that are themselves learned from the data. The hyperbolic tangent (tanh) is also used in the model to squeeze the information between $[-1,1]$ whenever additive structures are present. Figure \ref{fig:lstm_diag} gives a visual representation of the LSTM at time $t$. Note that at each time step, new information $z_t$ is added to the model and both the factors and cell state are updated accordingly. Note also that, given the recurrent structure of the LSTM in which different lags are processed sequentially by the internal functions, the number of parameters is not a function of the lag structure. This essentially means that a larger number of lags does not imply in more parameters. If time steps far back in time are not relevant for the prediction, the algorithm updates the parameters accordingly.

The equations governing the internal flow of information of the LSTM can be written as
\begin{equation}
    \begin{aligned}
        &f_{t|L}  = \phi_{t|L} \odot tanh (c_{t|L}) \\
        &c_{t|L}  = \psi_{t|L} \odot c_{t-1|L} + \zeta_{t|L} \odot tanh (w_{c}' z_t + u_{c}' f_{t-1|L} + b_{c}) \\
        &\phi_{t|L}  = sigmoid (w_{\phi}' z_t + u_{\phi}' f_{t-1|L} + b_{\phi}) \\
        &\psi_{t|L}  = sigmoid (w_{\psi}' z_t + u_{\psi}' f_{t-1|L} + b_{\psi}) \\
        &\zeta_{t|L}  = sigmoid (w_{\zeta}' z_t + u_{\zeta}' f_{t-1|L} + b_{\zeta}) \\
        &f_{0|L} = 0, \quad c_{0|L} = 0.
    \end{aligned}
\label{eq:lstm}
\end{equation}
Intuitively, the cell state $c_{t|L}$ is updated recursively where a first filter $\psi_{t|L}$ controls what \emph{past} information to retain, and a second filter $\zeta_{t|L}$ controls what \emph{new} information to retain. The factors $f_{t|L}$ are themselves a function of the cell state where a final filter $\phi_{t|L}$ controls what information from the cell state to use for prediction. 

\begin{figure}[h!]
    \centering
    \caption{The LSTM and FF-LSTM models}
    \begin{adjustbox}{max width=5\linewidth,center}
    \subfloat[\small The LSTM \vspace{-2mm} ]{{\includegraphics[width=0.8\textwidth]{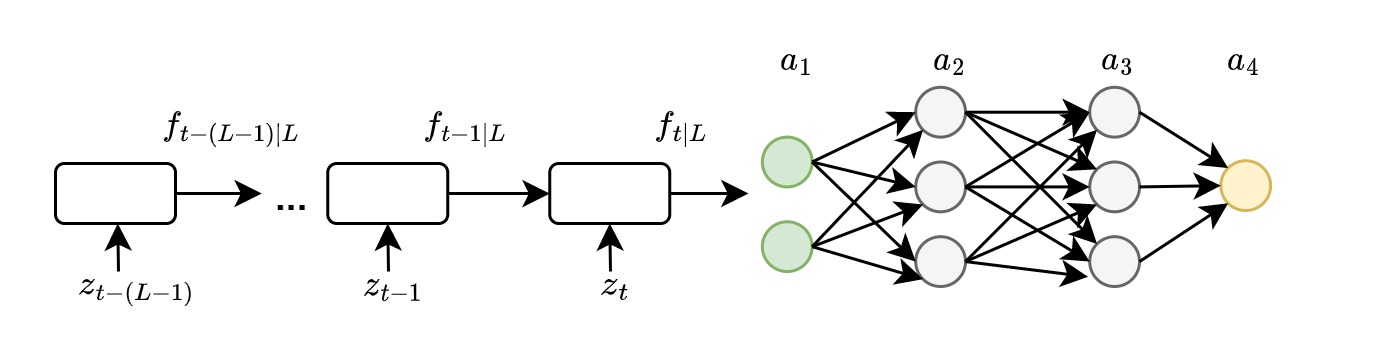}}}
    \end{adjustbox}
    \begin{adjustbox}{max width=5\linewidth,center}
    \subfloat[\small The FF-LSTM \vspace{-2mm}]{{\includegraphics[width=0.8\textwidth]{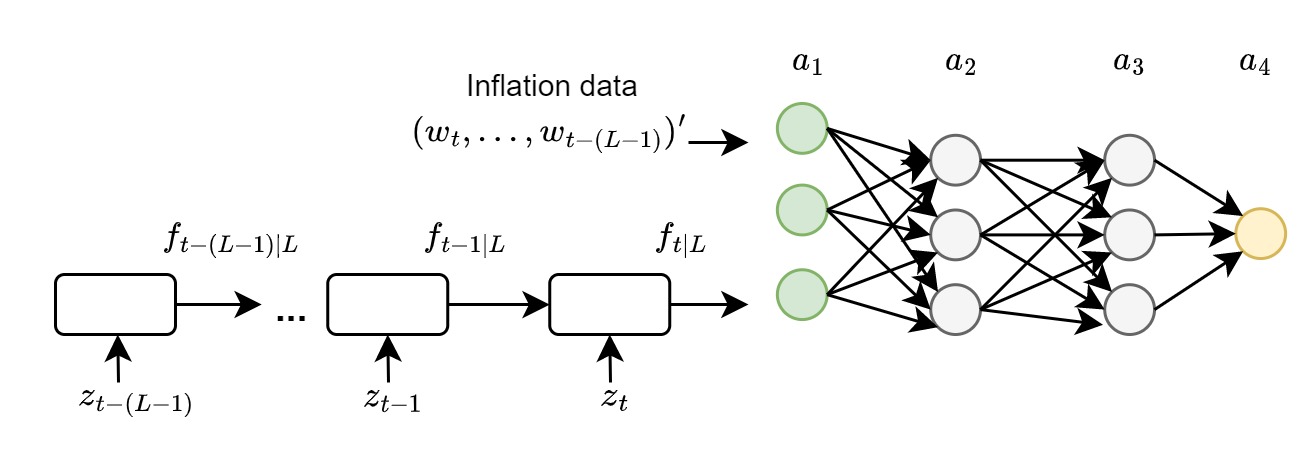}}}%
    \end{adjustbox}
    \caption*{\small Schematic representation of the models. $z_t$ is a vector collecting macroeconomic series except for price data, and $w_t$ is a vector collecting price data. $f_{t|L}$ collects the LSTM factors at time $t$. The feed-forward network is represented with $q+1=4$ layers and $n=3$ nodes per hidden layer.}
    \label{fig:lstm_fflstm}
\end{figure}

In this application, I stack a fully connected FF to the LSTM structure to get the final prediction following common practice. Hence, the prediction of the model is given by $G(x_t; \Theta_h) = g_{FF} \Big( f_{t|L} \Big)$, with $g_{FF}$ as defined in (\ref{eq:ff}) and $f_{t|L}$ as in (\ref{eq:lstm}). A graphical representation of the full LSTM model is in Figure \ref{fig:lstm_fflstm}.

The parameters of the model are (i) those from the LSTM structure, $(w_{(j)}, u_{(j)}, b_{(j)})'$ for $j=c,\phi,\psi,\zeta$, where $w_{(j)} \in \mathbb{R}^{N \times p}$,  $u_{(j)} \in \mathbb{R}^{p \times p}$, and $b_{(j)} \in \mathbb{R}^{p \times 1}$, as well as (ii) those from the FF network, $\{w_i,b_i\}_{i=1}^q$. The functions sigmoid and tanh are applied element-wise, and $\odot$ denotes the element-wise multiplication of two vectors. The hyperparameters are the number of LSTM factors $p$, the number of lags $L$, the number of nodes per layer $n$ in the FF network as well as the number of layers $q$. 

I consider estimating this model with two data sets: (i) the pool of economic predictors excluding inflation data, $x_t = (z_t,...,z_{t-(L-1)})'$, referred as LSTM-pool, and (ii) the complete data set, $x_t = (z_t,...,z_{t-(L-1)},w_t,...,w_{t-(L-1)})'$, referred as LSTM-all. The total number of parameters of the model amounts to $4(Np + p^2 + p) + (p+1)n +(Q-1)(n+1)n + (n+1)$.

\subsection{The FF-LSTM model}

The third model considered in this paper is constructed based on the two models described above. It can be viewed as an augmented FF-cpi model, where inflation data is combined to LSTM factors computed from the pool of macroeconomic predictors to form a composite input to a FF neural net. This model can be loosely associated with a factor-augmented distributed lag model, as it combines both types of inputs.

For clarification purposes, I rename the predictor sets as $x_t^w = (w_t,...,w_{t-(L-1)})'$ and $x_t^z = (z_t,...,z_{t-(L-1)})'$, and write $f_{t|L}(x_t^z)$ as the LSTM factors (here they are a function of the pool of macroeconomic predictors excluding price data). The FF-LSTM can then be defined as $G(x_t;\Theta_h) = g_{FF} \Big( (x_t^w,f_{t|L}(x_t^z))' \Big)$, where $g_{FF}$ is the FF network from (\ref{eq:ff}), $f_{t|L}$ is as in (\ref{eq:lstm}), and $\Theta_h$ collects the parameters of $f_{t|L}$ as well as the FF network. Similarly to the LSTM model, the hyperparameters are the number of LSTM factors $p$, the number of lags $L$, the number of nodes per layer $n$ in the FF network as well as the number of layers $q$. In this setting, the number of lags $L$ is not restricted to be the same across predictor sets $x_t^z$ and $x_t^w$. The number of parameters in this case amounts to $4(N^z p + p^2 + p) + (p+N^w L+1)n +(Q-1)(n+1)n + (n+1)$, where I denote $N^z$ and $N^w$ to be the number of predictor series in $x_t^z$ and $x_t^w$ respectively. Figure \ref{fig:lstm_fflstm} provides a graphical illustration.

\section{On the selection of hyperparameters}
\label{sec:estimation}

A central part in the estimation of neural networks is the selection of hyperparameters, usually implemented via cross-validation. For a FF model, this could involve selecting the number of layers and the number of nodes per layer in the network, for instance. Given an optimal set of hyperparameters, out-of-sample performance is then computed using an additional sample previously unseen by the model. In this paper, I use a modified version of the traditional \emph{k-fold} cross-validation that accounts for the autocorrelation inherent of time series. A more detailed description of the cross-validation procedure is available in Appendix \ref{app:model_spec}.

Table \ref{table:hyper} indicates the hyperparameters considered for cross-validation in this application, as well as the candidate and optimal values. For instance, the number of lags $L$ of predictors to be included in the model is a hyperparameter for which I select within the set $\{6,12,24,48\}$. The data used to estimate and cross-validate the models ranges from March-1959 to July-2006 (see Figure \ref{fig:cross_val}). As I also report results on the out-of-sample period starting from May-1993, I repeat the cross-validation process using only data prior to May-1993 as well. However, the resulting optimal hyperparameters using this later sample are found to be very similar to the ones reported in Table \ref{table:hyper}. For the cases where the hyperparameters are not the same, results on forecast performance are not qualitatively different. For this reason, I opt to set the same hyperparameters as in Table \ref{table:hyper} for both out-of-sample periods considered in the analysis.

A closer examination of the selected hyperparameters reveals that the networks are quite large. For example, the number of nodes is selected to be the upper bound of the grid-search values $\{16,32,64,128\}$ for all models, that is $n=128$. This means that the networks have huge amounts of parameters, i.e. $700K+$ for the FF-pool, or even $50K+$ for the LSTM models. Despite the common empirical knowledge that these models are able to handle cases with a much larger number of parameters than observations, practitioners may be interested in comparing the performance of such big models (i.e. networks with a disproportionately large amount of parameters than observations) to smaller networks in which we fix \emph{ex-ante} specific hyperparameters without the need to cross-validate those. Obvious advantages of considering fixed, small networks are a simplified cross-validation step and smaller computational burden. In order to provide insights on these issues, I consider estimating an additional set of models that share the same selected hyperparameters as before, except for the number of nodes $n$, which is fixed to be 4 for all models (I set 4 as an arbitrary small number). The number of nodes $n$ has by far the highest weight in the total number of parameters, which is decreased by approximately $98\%$ across models when fixing $n=4$ (e.g. from 51,017 to 1,125 in the case of the LSTM-pool, and from 758,273 to 22,725 for the FF-pool; see Table \ref{table:hyper_small}). Additionally, fixing only one hyperparameter, instead of multiple, facilitates the comparison. For instance, if one would also fix the number of lags to be smaller, it would be unclear whether differences in forecast performance are due to model complexity or the lag structure. Section \ref{sec:forecasting} discusses the differences in out-of-sample performance between the networks specified by the cross-validation step and the smaller networks as just defined.

\section{Data} 
\label{sec:data}

The data used in the empirical study is collected from the FRED-MD data base, a compilation of monthly US data made available by \cite{fred}, and corresponds to the vintage of October 2019. This data set comprises 128 series with 730 observations each, spanning the period from January-1959 to October-2019. Table \ref{table:data} in Appendix \ref{app:data} provides the description of all the series in the data set as well as information on the data transformation. 

The series undergo a sequence of transformations before estimation. First, I transform the data following \cite{fred} to guarantee stationarity (column Tcode in Table \ref{table:data}). The only exceptions are aggregate CPI and aggregate PCE (items No. 110 and 120 respectively in Table \ref{table:data}), which are transformed in log differences. The inflation target in the paper is then defined as $\pi_t = log(P_t) - log(P_{t-1})$, where $P_t$ is either aggregate CPI or aggregate PCE at time $t$. Second, missing observations are replaced using the iterative expectation-maximization (EM) algorithm from \cite{stock02} (the percentage of missing values for each series is reported in Table \ref{table:data}). Finally, for the estimation of the neural network models, the data is further normalized within the interval $[-1,1]$. The normalization is carried out in-sample and extrapolated to the out-of-sample period to avoid look-ahead bias. This transformation is appropriate here since the evidence suggests that neural nets perform better when the regressors share a similar order of magnitude.\footnote{This fact is related to the gradient descent optimization process, which seems to converge much faster when the regressors are normalized (\citealp{norm}). For details on the optimization procedure, see appendix \ref{app:estimation}.} This scaling is similar to assuming an equal importance of covariates \emph{ex-ante}. 

After data transformations, I collect into $w_t$ the price series corresponding to items No. 110 to 119 when targeting CPI, or items No. 120 to 123 when targeting PCE, from Table \ref{table:data}. In contrast, $z_t$ corresponds to all macroeconomic series from Table \ref{table:data} that are not in $w_t$. For instance, when targeting CPI, $z_t$ includes all series except items No. 110 to 119.

\section{The LSTM factors}
\label{sec:comp}

The LSTM and FF-LSTM models work as a dimensionality-reduction tool, as they squeeze the input information from a potentially large set of variables into a $p$-dimensional vector $f_{t|L}$, collecting what is referred as LSTM factors. In this section, I compare these factors with those extracted from PCA as well as business cycle indicators. I show that the estimated factors, in addition to reducing the dimensionality, also convey information on variable selection relevant for prediction, as per the supervised nature of the estimation, as discussed previously.

I extract LSTM factors by evaluating the estimated models over the full sample period, and obtain $f_{t|L}$ for $t=L+h,...,T$.\footnote{The number of factors as well as other hyperparameters are selected via cross-validation according to Table \ref{table:hyper}.} However, as the estimation of neural networks can be sensitive to initial values, these models are generally estimated as an ensemble over neural networks embedding different initializations (see Section \ref{sec:ensemble} for more details). In order to recover a unique set of LSTM factors, I therefore estimate $K$ models $\{ \mathcal{M}_k \}_{k=1}^K$ with different initial values, and select the model $\mathcal{M}_k$ with lowest out-of-sample error over a validation set. This essentially implies selecting an optimal initialization.

\begingroup
\renewcommand{\arraystretch}{0.9}
\begin{table} \centering \footnotesize
\caption{Variance decomposition of LSTM factors}
\begin{threeparttable}
\begin{tabular}{ C{1.59cm} C{0.5cm} c R R R R R R R R R}
  \toprule
   & & &
  \multicolumn{1}{c} {\footnotesize Out\&Inc} &
  \multicolumn{1}{c} {\footnotesize Lab Mkt} &
  \multicolumn{1}{c} {\footnotesize Hous} &
  \multicolumn{1}{c} {\footnotesize Orders} &
  \multicolumn{1}{c} {\footnotesize Mon\&Cr} &
  \multicolumn{1}{c} {\footnotesize Stock Mkt} &
  \multicolumn{1}{c} {\footnotesize IR\&ER} &
  \multicolumn{1}{c} {\footnotesize Prices} & \multicolumn{1}{c} {\footnotesize CPI/PCE} \\ 
  \multicolumn{12}{l}{\textit{\footnotesize Principal components}} \\ \hline
    \multicolumn{3}{c}{\footnotesize PC1} & 0.35  & 0.26  & 0.31  & 0.16  & 0.02  & 0.07  & 0.06  & 0.00  & 0.02 \\
    \multicolumn{3}{c}{\footnotesize PC2} & 0.03  & 0.02  & 0.13  & 0.09  & 0.01  & 0.06  & 0.19  & 0.03  & 0.47 \\
    \multicolumn{12}{l}{\textit{\footnotesize Neural network models}}  \\ \hline
    LSTM-pool & 3     & f1    & 0.01  & 0.04  & 0.15  & 0.06  & 0.01  & 0.01  & 0.07  & 0.01  & 0.21 \\
          &       & f2    & 0.01  & 0.01  & 0.01  & 0.07  & 0.01  & 0.01  & 0.16  & 0.01  & 0.33 \\
          & 6     & f1    & 0.05  & 0.05  & 0.16  & 0.02  & 0.01  & 0.02  & 0.04  & 0.00  & 0.21 \\
          &       & f2    & 0.01  & 0.03  & 0.28  & 0.04  & 0.01  & 0.00  & 0.07  & 0.00  & 0.25 \\
          & 12    & f1    & 0.00  & 0.01  & 0.01  & 0.04  & 0.01  & 0.00  & 0.05  & 0.01  & 0.20 \\
          &       & f2    & 0.02  & 0.04  & 0.11  & 0.09  & 0.01  & 0.02  & 0.11  & 0.01  & 0.27 \\
          & 24    & f1    & 0.00  & 0.02  & 0.20  & 0.03  & 0.01  & 0.00  & 0.07  & 0.00  & 0.13 \\
          &       & f2    & 0.00  & 0.04  & 0.04  & 0.03  & 0.01  & 0.01  & 0.08  & 0.00  & 0.31 \\
    LSTM-all & 3     & f1    & 0.00  & 0.02  & 0.09  & 0.05  & 0.01  & 0.00  & 0.07  & 0.01  & 0.38 \\
          &       & f2    & 0.03  & 0.05  & 0.03  & 0.09  & 0.01  & 0.01  & 0.10  & 0.01  & 0.24 \\
          & 6     & f1    & 0.01  & 0.02  & 0.05  & 0.07  & 0.01  & 0.04  & 0.19  & 0.01  & 0.45 \\
          &       & f2    & 0.00  & 0.03  & 0.03  & 0.03  & 0.01  & 0.01  & 0.05  & 0.00  & 0.20 \\
          & 12    & f1    & 0.02  & 0.04  & 0.35  & 0.03  & 0.01  & 0.00  & 0.01  & 0.00  & 0.06 \\
          &       & f2    & 0.05  & 0.11  & 0.08  & 0.06  & 0.01  & 0.02  & 0.05  & 0.01  & 0.00 \\
          & 24    & f1    & 0.08  & 0.07  & 0.39  & 0.02  & 0.01  & 0.01  & 0.02  & 0.01  & 0.01 \\
          &       & f2    & 0.00  & 0.02  & 0.08  & 0.06  & 0.01  & 0.02  & 0.10  & 0.02  & 0.32 \\
    FF-LSTM & 3     & f1    & 0.01  & 0.02  & 0.11  & 0.09  & 0.01  & 0.02  & 0.11  & 0.00  & 0.30 \\
          &       & f2    & 0.04  & 0.09  & 0.19  & 0.09  & 0.01  & 0.00  & 0.06  & 0.00  & 0.20 \\
          & 6     & f1    & 0.02  & 0.02  & 0.05  & 0.06  & 0.01  & 0.08  & 0.16  & 0.01  & 0.34 \\
          &       & f2    & 0.09  & 0.10  & 0.47  & 0.07  & 0.01  & 0.01  & 0.04  & 0.00  & 0.09 \\
          & 12    & f1    & 0.04  & 0.06  & 0.39  & 0.07  & 0.01  & 0.00  & 0.03  & 0.00  & 0.10 \\
          &       & f2    & 0.04  & 0.08  & 0.19  & 0.10  & 0.01  & 0.01  & 0.11  & 0.00  & 0.09 \\
          & 24    & f1    & 0.13  & 0.16  & 0.18  & 0.04  & 0.01  & 0.05  & 0.04  & 0.00  & 0.01 \\
          &       & f2    & 0.04  & 0.05  & 0.45  & 0.03  & 0.01  & 0.00  & 0.02  & 0.00  & 0.01 \\
  \bottomrule
\end{tabular}
\begin{tablenotes}
\small
\item This table presents the average variance decomposition of the estimated LSTM factors (and the first two PCA factors) across groups of variables (groups as defined in \citealp{fred}). The entries correspond to the squared correlation between the factors and each variable in the original data set $\mathcal{Z}$, averaged across groups. `Prices' excludes CPI and PCE, which are shown in the last column separately. Results are reported for the $p=2$ LSTM factors at horizons $h=3,6,12,24$. Darker shades of grey correspond to higher values.
\vspace{2mm}
\end{tablenotes}
\end{threeparttable}
\label{table:heatmap}
\end{table}
\endgroup

To shed light on the information encoded into the LSTM factors, Table \ref{table:heatmap} shows how these factors relate to each group of variables from the set of predictors $\mathcal{Z}$, and offers a comparison with standard PCA factors. The entries correspond to the squared correlation between the factors and each variable in $\mathcal{Z}$, averaged across groups. Three points stand out. First, LSTM factors are generally loaded on common predictors of inflation, e.g. prices, interest \& exchange rates, housing starts, employment. Second, apart from the anticipated high correlation with CPI and PCE variables, LSTM factors are significantly loaded on housing starts. This group was previously reported in the literature as important for predicting inflation (see for instance \citealp{stock99} and \citealp{bai_ng}). Third, FF-LSTM factors tend to be \emph{more} correlated with real variables (housing, output and labour market) and \emph{less} correlated with price variables than the LSTM-pool or LSTM-all factors, especially at large horizons. Recall that the factors extracted from the LSTM-pool and the FF-LSTM both receive the pool of macroeconomic predictors excluding price data as inputs, but that the FF-LSTM observes price information separately that it uses to improve the forecast. The fact that FF-LSTM factors give relatively less weight to price information and relatively more weight to real variables suggests that these factors work as filters: They filter out from the pool of macro predictors the signals to forecast inflation that are less related to prices, since the later is already being provided to the model separately. I borrow from these insights and interpret FF-LSTM factors as ``output-driven'' and LSTM-pool/LSTM-all factors as ``price-driven'' in the visual comparison with macroeconomic indicators below.

\begin{figure}
\caption{Individual variable importance in LSTM factors}
    \centering
    \includegraphics[width=10cm]{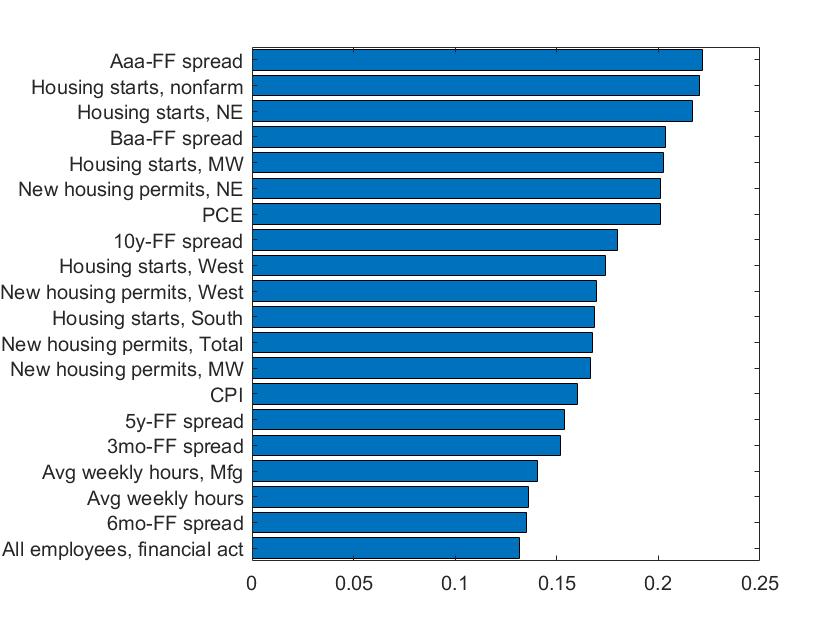}
    \label{fig:vardecomp}
\caption*{\small The entries correspond to the squared correlation between LSTM factors and each variable in $\mathcal{Z}$, averaged across models (LSTM-pool, LSTM-all, FF-LSTM), fators (f1 and f2), and horizons (3, 6, 12, 24). Presented are the 20 variables with higher loadings.}
\end{figure}

To dig deeper into variable importance embedded into the LSTM factors, Figure \ref{fig:vardecomp} shows the $20$ variables in $\mathcal{Z}$ presenting the higher loadings. The entries are the same as in Table \ref{table:heatmap}, i.e. the squared correlation between the factors and the variables, only this time they are averaged across models (LSTM-pool, LSTM-all, FF-LSTM), factors (f1 and f2), and horizons (3, 6, 12, 24). High-yield corporate bond spreads stand out, together with housing starts and permits. PCE and CPI can be categorized as the third most important group, followed by labor market variables, with 20+ variables appearing right below in the ranking (not reported).

\begin{figure}[h!]
\caption{The LSTM factors}
\centering
\begin{adjustbox}{max width=5\linewidth,center}
\subfloat[\small Price-driven factors \vspace{-2mm} ]{{\includegraphics[width=1.2\textwidth]{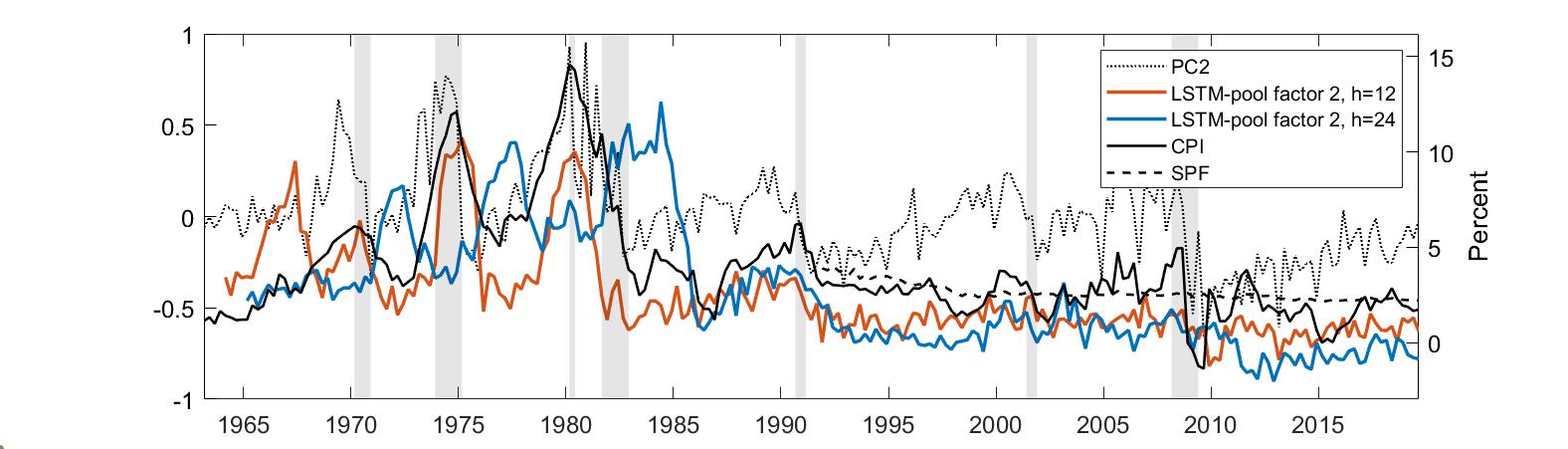}}}
\end{adjustbox}
\begin{adjustbox}{max width=5\linewidth,center}
\subfloat[\small Output-driven factors \vspace{-2mm}]{{\includegraphics[width=1.2\textwidth]{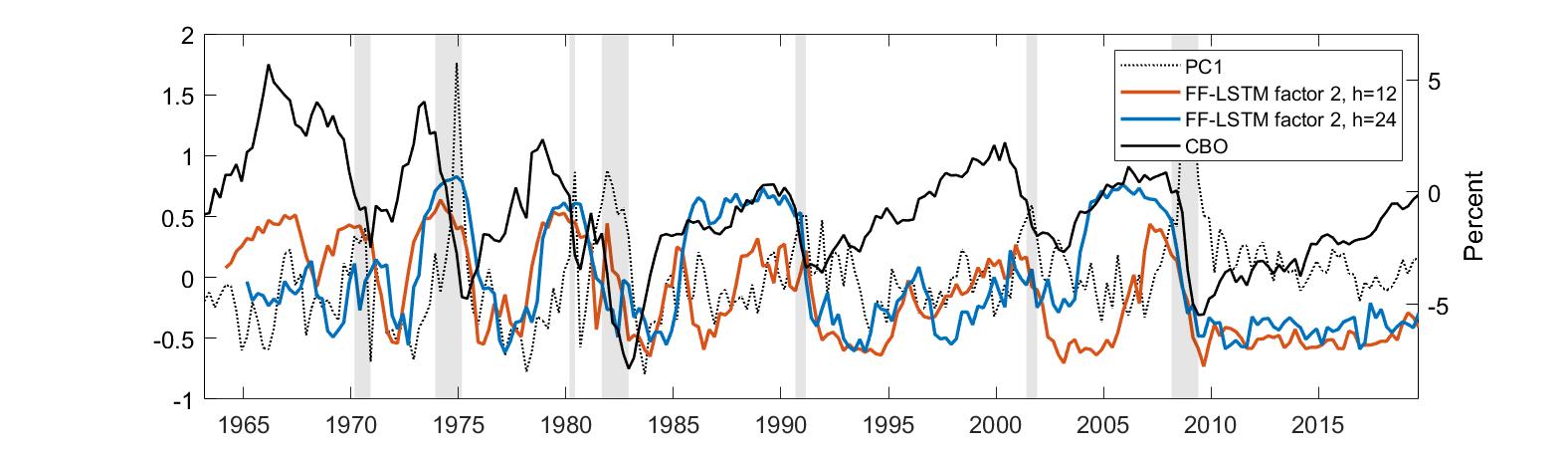}}}%
\end{adjustbox}
\caption*{\small The figure plots selected LSTM factors estimated from the LSTM-pool (subplot (a)) and FF-LSTM (subplot (b)) over the full sample period $t=L+h,...,T$, at horizons 12 and 24. The forecasting target is CPI inflation. For comparison, the plots also show the first two principal components (PC1 and PC2) estimated form the input data $\mathcal{Z}$, as well as headline CPI inflation, 10-year CPI inflation according to the Survey of Professional Forecasters (SPF), and the output gap from the Congressional Budget Office (CBO). The grey areas are NBER recessions. The LSTM factors and principal components are plot on the left axis (series are normalized), and the other series on the right axis (percent).}
\label{fig:factors}
\end{figure}

Looking at the dynamics, Figure \ref{fig:factors} plots selected factors over time for horizons $h=12,24$. In subplot (a), note how price-driven factors are correlated with headline inflation ($0.73$ at $h=12$, and $0.56$ at $h=24$). Compared to PC2, the principal component more correlated with prices (see Table \ref{table:heatmap}), LSTM factors seem to better capture the downward trend of inflation in the second part of the sample, and are less volatile. The plot also shows the long-term trend of CPI inflation according to the Survey of Professional Forecasters (SPF) for comparison. This may indicate that the LSTM can be useful at detecting low frequency components.

\vspace{2mm}

Subplot (b) shows factors from the FF-LSTM model and highlights its similarity with the business cycle. I compare the LSTM factors with the output gap produced by the Congressional Budget Office (CBO), NBER recessions and PC1. Interestingly, note how the factors lag the CBO's output gap measure while being relatively correlated with the later ($0.53$ at $h=12$, and $0.45$ at $h=24$). This again speaks to the ability of the LSTM to condense macroeconomic data in an economic-meaningful way, as the model seems to successfully detect business cycle fluctuations. Additionally, given that LSTM factors are estimated within a prediction context, these findings are consistent with Phillips curve-based models, since output gap-type signals appear as an important inflation predictor themselves (other recent works emphasizing a revival of Phillips curve methods are e.g. \citealp{gio} and \citealp{coulombePC}). It is interesting to note that during the post-Global Financial Crisis (GFC) period, the FF-LSTM factors remain at its lowest historic levels and are more pessimistic than the CBO's measure of the output gap. This may be an indication that the \emph{de facto} recovery may have been more sluggish than previously thought. 

It is worth mentioning that the small number of LSTM factors used above is a direct consequence of the cross-validation design, in which the number of factors is restricted to be within the set $\{2,4,6,8\}$ (see Table \ref{table:hyper} on hyperparameter selection). This is a modelling choice that facilitates interpretation and comparison with PCA and other factor structures usually encountered in the empirical literature. For instance, \cite{fred} report the optimal number of factors estimated from the FRED-MD database to be between 2 and 8 in the context of a standard factor model, using a similar sample period than the present analysis. For the practitioner mainly concerned with forecast performance, it may be advisable to investigate a finer grid of number of factors.

\section{Forecast performance}
\label{sec:forecasting}

This section compares the forecast performance of neural networks with common benchmarks from the literature of inflation forecasting and alternative machine learning models. These comprise the autoregressive model of order $p$, or $AR(p)$, the factor model (Fact), the unobserved components with stochastic volatility model (UCSV), the random forest (RF), as well as shrinkage methods, LASSO, Ridge regression and the Elastic-net (ENet). The description of the benchmarks is in Appendix \ref{app:bench}.

Forecasts are computed in a rolling window basis. At each time step, a window of $R$ observations is used to estimate the models (and to cross-validate hyperparameters), and forecast performance is averaged over $P$ out-of-sample (OOS) observations. The total number of observations in the sample is equal to $T = R+P+h-1$, where $h$ is the forecast horizon. Note that $R$ changes with the forecast horizon, as $T$ and $P$ are fixed. Following standard practice, I evaluate forecast performance using the Root Mean Squered Error (RMSE),
\begin{equation*}
    RMSE_{h,m} = \sqrt{\frac{1}{P} \mathlarger{\mathlarger{\sum}}_{t \in OOS} \left( \pi_t - \hat{\pi}_{t,m} \right)^2},
\end{equation*}
where $\hat{\pi}_{t,m}$ is the prediction of model $m$ at time $t$. To compare models, I employ a standard \cite{diebold} (DM) test with respect to the $AR(p)$, as well as the Model Confidence Set (MCS) proposed in \cite{hansen_mcs}. Given an initial set of predictive models, the MCS is the subset (potentially equal to the original) that contains the best model at a given level of confidence. I use $5000$ bootstrap samples to construct the MCS test statistics, and compute $75\%$ model confidence sets following the original paper.

For the neural networks, I re-estimate the models at every 4 years during the OOS period.\footnote{I opted for this solution to improve computational time, although a more frequent re-estimation of the neural network models, e.g. every year, may improve performance.} Results are presented for two distinct OOS periods, (i) May-1993 to July-2006, and (ii) August-2006 to October-2019, the second period covering the GFC and featuring more volatile inflation. Results refer to models with CPI inflation as the main target (similar analysis is done for PCE inflation in Appendix \ref{app:OFS_app}).

\subsection{Predictions as an ensemble}
\label{sec:ensemble}

The usual non-convex loss function of neural networks implies that the estimated parameters are in general very sensitive to initial values.\footnote{Initial parameters are randomly drawn from a specific uniform distribution (Appendix \ref{app:estimation}).} In practice, this means that the neural network predictions will be very much dependent on the initialization. To overcome this issue, the empirical literature usually adopts the solution of averaging out the predictions from models estimated with different initial values (see e.g. \citealp{izmailov2019}, \citealp{fort2020}, and \citealp{wilson2022} for papers exploring model averaging in the context of deep neural nets).

Consider the set of models $\{ \mathcal{M}_k \}_{k=1}^K$ estimated from the same neural network architecture but with different initial values. Let $\{ \hat{y}_{k,t+h} \}_{t=1}^P$ be the forecast associated with model $\mathcal{M}_k$, where $P$ is the OOS size. The ensemble prediction at period $t+h$ is defined as $\hat{y}_{ens,t+h} = \frac{1}{K} \sum_{k=1}^K \hat{y}_{k,t+h}$, which is essentially a model-averaging technique that attributes an equal weight to each forecast $\hat{y}_{k,t+h}$. Section \ref{sec:iv} shows empirically that this solution is at least as good as the individual forecasts in terms of mean squared error according to a standard DM test procedure.

I therefore adopt this ensemble technique and estimate each neural network $K=1400$ times letting vary the initialization.\footnote{In order to guarantee optimal computational time, the number of iterations was chosen such that it is proportional to the number of available processors.} The main OOS results reported below refer to the ensemble prediction $\hat{y}_{ens,t+h}$ across the $1400$ forecasts.

\subsection{Out-of-sample results}
\label{sec:OFS}

Table \ref{table:rmse} presents the OOS results. The entries are loss ratios with respect to the $AR(p)$ model. A glance at Table \ref{table:rmse} reveals that neural networks performance is competitive against benchmarks, although not outstanding. Specifically, neural nets can present substantial forecasting gains (e.g. RMSE ratios $\approx 0.85$), but these are not robust across models and horizons. LSTM-only models (LSTM-pool, LSTM-all) in particular tend to present a satisfactory forecast performance at long horizons as well as at horizon 1. On the other hand, the performance of FF models (FF-cpi, FF-pool, FF-LSTM) changes with the sample and the input data. I analyse below the local forecasting performance, as opposed to the average performance, to shed more light on these points. It is interesting to note that adding information on the state of the economy usually helps in forecast performance (FF-LSTM model as opposed to the FF-cpi model). Conversely, adding price information when the pool of macro predictors is already taken into account does not seem to improve the prediction significantly (LSTM-all as opposed to LSTM-pool). It is worth mentioning that the random forest stands out overall across the different scenarios and forecast horizons, a finding that is in line with previous literature (\citealp{medeiros}). Also note that the performance of the neural nets and some benchmarks improves in the 2006-2019 period with respect to 1993-2006 (especially when predicting PCE inflation; see Table \ref{table:rmse_pce}). This fact resonates the common understanding that machine learning models tend to outperform traditional benchmarks during the post-GFC period of heightened macroeconomic uncertainty and more volatile inflation (see for instance \citealp{medeiros} and \citealp{buckmann2022} for applications to inflation and unemployment respectively).

\begin{table}[h!] \centering \footnotesize
\begin{threeparttable}
  \caption{Out-of sample forecast performance for CPI inflation}
   \begin{tabular}{lccccccccccc}
   \toprule
          & \multicolumn{5}{c}{1993-2006}         &       & \multicolumn{5}{c}{2006-2019} \\
\cmidrule{2-6}\cmidrule{8-12}          & 1     & 3     & 6     & 12    & 24    &       & 1     & 3     & 6     & 12    & 24 \\
    \multicolumn{6}{l}{\textit{Benchmark models}} &       &       &       &       &       &  \\
    \midrule
    \quad UCSV  & 1.02  & \textbf{1.07} & 1.00  & 1.07  & \textbf{0.91**} &       & \textbf{1.08} & 0.99  & \textbf{0.98} & \textbf{1.02} & \textbf{1.02} \\
    \quad Fact  & 1.00  & \textbf{1.00} & 1.01  & 1.04  & \textbf{0.99} &       & \textbf{0.95} & \textbf{1.00} & \textbf{1.01} & \textbf{1.03} & \textbf{1.02} \\
    \quad RF    & \textbf{0.88***} & \textbf{0.98} & \textbf{0.96*} & \textbf{0.92***} & \textbf{0.92*} &       & \textbf{0.99} & \textbf{0.89***} & \textbf{0.90***} & \textbf{0.94***} & \textbf{0.94***} \\
    \quad LASSO  & \textbf{0.91***} & \textbf{0.96} & \textbf{0.97**} & 1.01  & \textbf{1.01} &       & \textbf{0.95***} & \textbf{0.94**} & \textbf{0.95***} & \textbf{0.94***} & \textbf{0.91***} \\
    \quad Ridge & 0.94** & 1.01  & 1.02  & \textbf{0.98} & \textbf{0.93***} &       & \textbf{1.03} & 1.15  & \textbf{1.35} & 1.23  & 1.36 \\
    \quad ENet & \textbf{0.91***} & \textbf{0.96*} & \textbf{0.97**} & 1.01  & \textbf{1.01} &       & \textbf{0.93***} & \textbf{0.93***} & \textbf{0.95***} & \textbf{0.93***} & \textbf{0.90***} \vspace{2mm} \\
    \multicolumn{6}{l}{\textit{Neural network models}} &       &       &       &       &       &  \\
    \midrule
    \multicolumn{6}{l}{\textit{Large architecture}} &       &       &       &       &       &  \\
    \quad FF-cpi & \textbf{0.85***} & \textbf{0.92**} & \textbf{0.91***} & \textbf{0.98} & \textbf{1.07} &       & \textbf{1.04} & \textbf{0.97} & \textbf{0.98} & \textbf{1.01} & \textbf{0.94***} \\
    \quad FF-pool & 0.99  & \textbf{0.97} & \textbf{0.96*} & \textbf{0.93***} & \textbf{0.94*} &       & \textbf{1.05} & \textbf{0.85***} & \textbf{0.87***} & \textbf{0.95} & \textbf{0.99} \\
    \quad LSTM-pool & \textbf{0.85***} & 1.08  & 1.08  & \textbf{0.93**} & \textbf{0.98} &       & \textbf{0.94*} & \textbf{0.88**} & \textbf{1.02} & \textbf{0.95*} & \textbf{0.93***} \\
    \quad LSTM-all & \textbf{0.84***} & 1.08  & 1.08  & \textbf{0.93*} & \textbf{0.97} &       & \textbf{0.93**} & \textbf{0.88**} & \textbf{1.03} & \textbf{0.95} & \textbf{0.94**} \\
    \quad FF-LSTM & \textbf{0.85***} & \textbf{0.92*} & \textbf{0.92***} & \textbf{0.97*} & \textbf{1.05} &       & \textbf{1.02} & \textbf{0.94*} & \textbf{0.96} & \textbf{0.98} & \textbf{0.92***} \\
    \multicolumn{6}{l}{\textit{Small architecture}} &       &       &       &       &       &  \\
    \quad FF-cpi & \textbf{0.91*} & \textbf{0.96} & \textbf{0.96} & 1.02  & 1.06  &       & 1.07  & \textbf{0.97} & \textbf{0.97} & \textbf{1.01} & \textbf{0.99} \\
    \quad FF-pool & 1.00  & 1.04  & 1.01  & \textbf{0.95**} & \textbf{0.91***} &       & \textbf{1.06} & \textbf{0.88**} & \textbf{0.90**} & \textbf{0.98} & \textbf{1.02} \\
    \quad LSTM-pool & \textbf{0.89**} & 1.12  & 1.11  & \textbf{0.95*} & \textbf{0.97} &       & \textbf{1.00} & \textbf{0.87***} & \textbf{0.89**} & \textbf{0.94***} & \textbf{0.96**} \\
    \quad LSTM-all & \textbf{0.89**} & 1.11  & 1.11  & \textbf{0.95*} & \textbf{0.94*} &       & \textbf{0.99} & \textbf{0.88***} & \textbf{0.89**} & \textbf{0.94**} & \textbf{0.96**} \\
    \quad FF-LSTM & \textbf{0.88**} & \textbf{0.98} & \textbf{0.96} & \textbf{0.96*} & \textbf{0.95*} &       & \textbf{1.01} & \textbf{0.91**} & \textbf{0.92**} & \textbf{0.97} & \textbf{0.95**} \\
    \bottomrule
    \end{tabular}%
    \begin{tablenotes}
    \small
    \item The table presents the loss ratios with respect to the $AR(p)$ model for horizons $h=1,3,6,12,24$ and two OOS periods. The target is CPI inflation. \emph{Large architecture} and \emph{Small architecture} refer to models specified as in table \ref{table:hyper} and \ref{table:hyper_small} respectively. The loss function is the RMSE. *, **, *** denote significance of the one-sided DM test at a $10\%$, $5\%$ and $1\%$ levels respectively. Models retained in the $75\%$ MCS are in bold.
    \vspace{2mm}
    \end{tablenotes}
    \label{table:rmse}
\end{threeparttable}
\end{table}%

On the comparison between large and small networks, the former specified by cross-validation and the later by networks with $4$ nodes per layer, I discuss a few takeaways. First, the performance of small networks is in general worst (in the sense of DM tests) than large ones, but there are exceptions. Specifically, smaller networks appear to help the performance of LSTM models at medium horizons (6 in particular) for the 2006-2019 sample (note that this gain in accuracy comes entirely from the FF network stacked to the LSTM). Second, forecasting gains of smaller networks remain substantial. This suggests that overly complex networks do not necessarily guarantee a better performance. These same insights follow for PCE (Table \ref{table:rmse_pce}). In general, these results speak in favor of ensemble techniques that take into account not a single architecture but multiple if the interest remains solely in forecast performance. Regarding computational time, despite the large difference in the number of parameters in this application, small networks were on average only $30\%$ faster than larger ones, which doesn't seem to compromise the use of the latter within cross-validation or ensemble model designs.

\begin{figure}[h!]
\caption{Local forecast accuracy and Predictions for CPI inflation}
\centering
\begin{adjustbox}{max width=5\linewidth,center}
\subfloat[\small Out-of-sample 1993-2006  \vspace{-2mm} ]{{\includegraphics[width=1.05\textwidth]{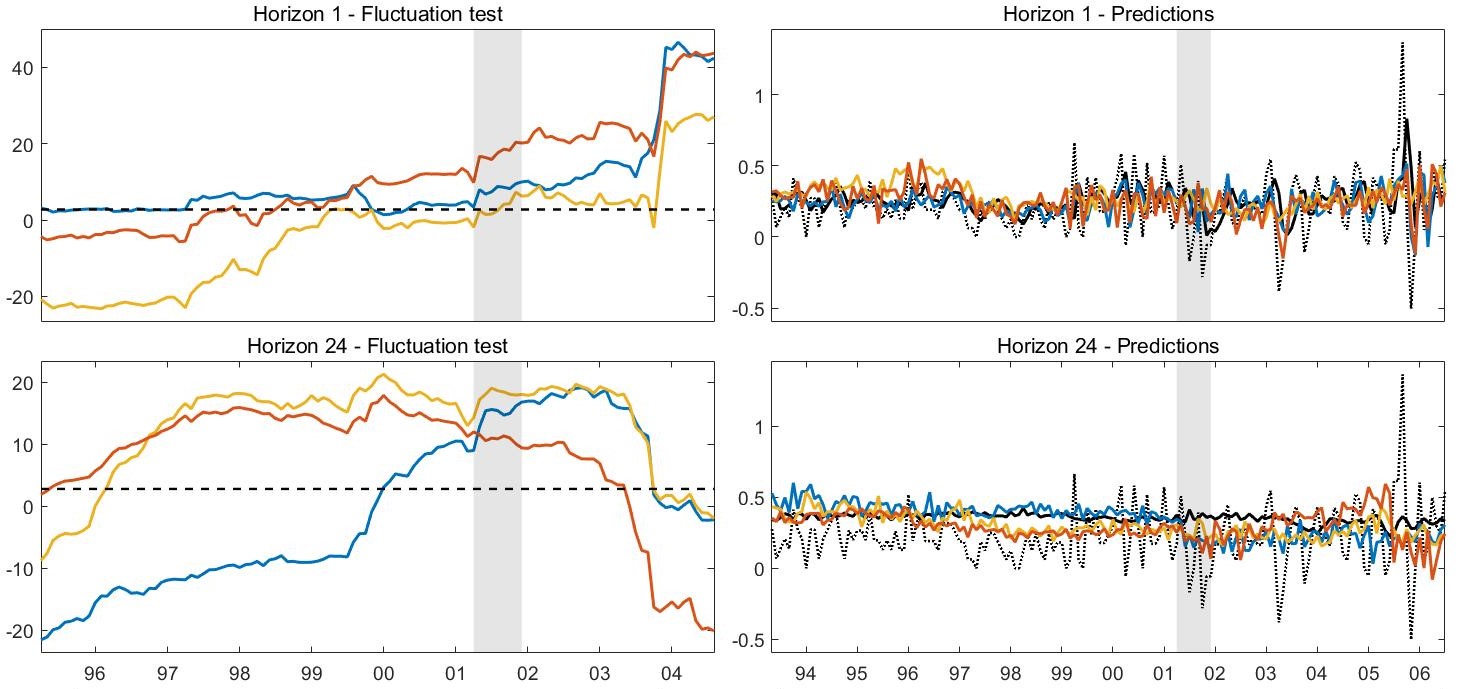}}}
\end{adjustbox}
\begin{adjustbox}{max width=5\linewidth,center}
\subfloat[\small Out-of-sample 2006-2019 \vspace{-2mm}]{{\includegraphics[width=1.05\textwidth]{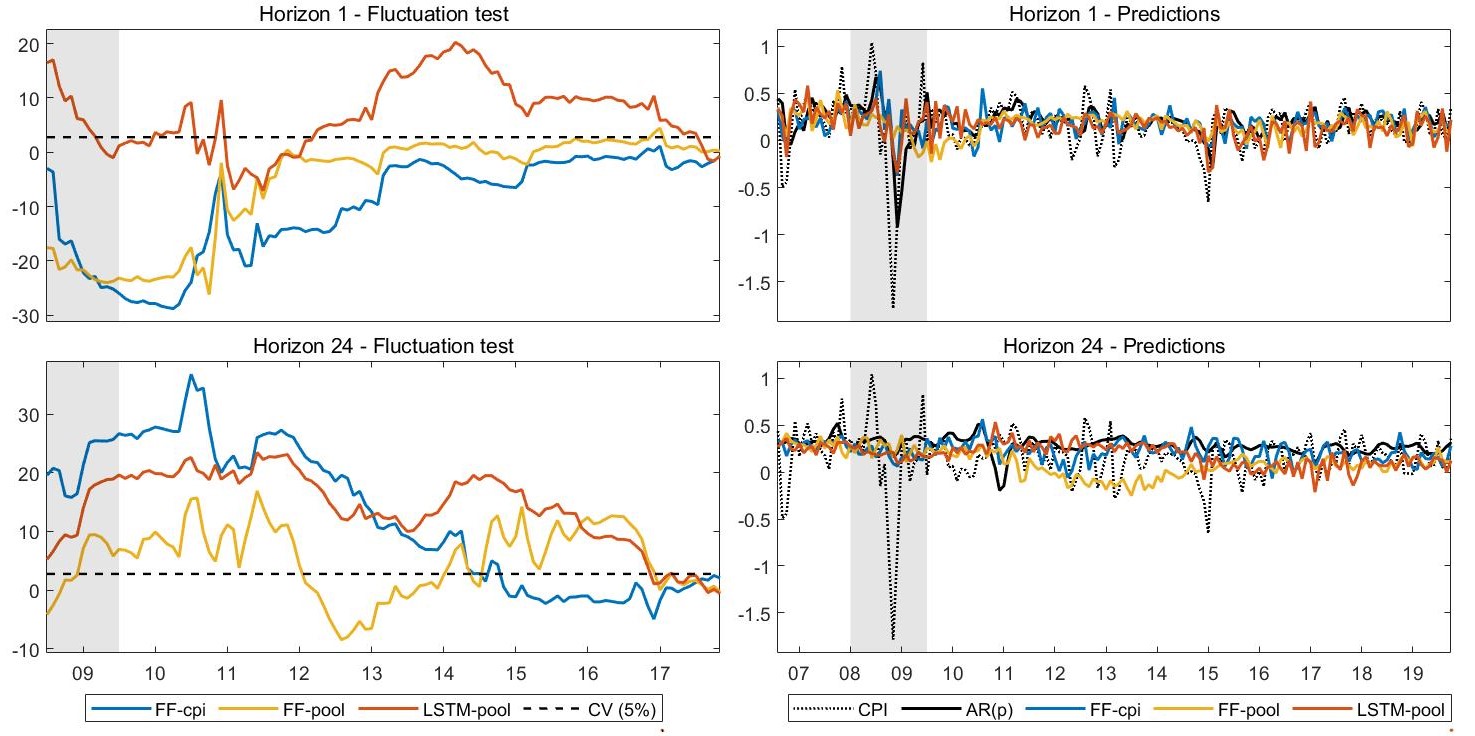}}}%
\end{adjustbox}
\caption*{\small First column: Test statistics according to the Fluctuation Test from \cite{rossi}; values above the dashed line indicate that the underlying model presents better forecast accuracy than the $AR(p)$ benchmark over a window of fixed size at the $5\%$ confidence level. Second column: Predictions of selected models; the scale is $100 \times$ log differences. Gray areas are NBER recessions.}
\label{fig:ft}
\end{figure}

In order to get a clearer picture of the models' performance over time, I employ the Fluctuation Test introduced in \cite{rossi}. The proposed statistic tests for an equal forecast accuracy of two different forecast series and is robust to instability on their relative forecast performance. The statistic is much similar to the DM test procedure, except that it is computed over a rolling window of fixed size $m=48$. In this application, I use the one-sided test where the rejection of the null hypothesis (i.e. a test statistic larger than the critical value) implies that the candidate forecast series of each neural net is statistically superior to the $AR(p)$ benchmark for a given observation in the OOS period. The value of $m$ is chosen such that $m/P \approx 0.3$, in which case the one-sided critical value at the $5\%$ confidence level is $2.77$ according to \cite{rossi}. Figure \ref{fig:ft} shows the implied test statistics over time at selected horizons with the CPI as main target, and focuses on the case of large architectures, i.e. models selected via cross-validation. (In the plots, I do not include the LSTM-all and FF-LSTM models as their overall performances are very similar to the LSTM-pool and FF-cpi models respectively, and I henceforth refer to the LSTM-pool as simply LSTM in this section). 

I discuss three points. First, the LSTM model seems suitable to forecast during periods of higher macroeconomic uncertainty as was the case in the post-GFC period (2006-2019). Note how the LSTM is superior to the $AR(p)$ (as well as the FF models to some extent) for horizons 1 and 24 (Figure \ref{fig:ft}) and horizons 3 and 6 (Figure \ref{fig:ft_app} in the appendix). The prediction plots reveal that this is mainly because the LSTM is on average less affected by sudden, short-lived movements in prices compared to other models. For instance, some models are very sensitive to the downward pressure on prices caused by the GFC, as is the case of the $AR(p)$, but also the random forest (plot not reported).

Second, the LSTM appears to be a good model at long horizons, a finding that is robust to the OOS period and network architecture, while forecasting gains can be found at very short horizons as well, depending on the sample. Results for medium horizons are somewhat mixed, as it generally depends on the architecture. More specifically, the prediction plots at horizon 1 show that the LSTM adapts quite well to short term movements in inflation, e.g. note in particular the second half of the 1993-2006 sample. However this is not a particularity of the LSTM, as the FF-cpi and FF-LSTM models are as good. At the longest horizon, the LSTM maintains a good forecast accuracy during the majority of the sample in both OOS periods. The long-term predictions of the LSTM tend to be quite smooth and to match the direction of inflation 24 steps ahead, suggesting an ability to predict well the trend of the series. There is one exception to the good performance of the LSTM at horizon 24 in the end of the 1993-2006 sample, which drives the modest average result reported in Table \ref{table:rmse}. The prediction plot helps understand that episode, which is characterized by a large swing in headline inflation. The LSTM did particularly bad and the $AR(p)$ particularly well, where the latter predicts a stable trend. Looking at medium horizons, LSTM models do not appear to be good predictors in general, although the small networks show gains at the 2006-2019 sample, suggesting that the poor performance can be in fact related to the stacked FF architecture. It seems that at these horizons the LSTM predictions are somehow more ``erratic'' than those of FF and autoregressive models (Figure \ref{fig:ft_app}), which on average contributes negatively to its performance. In fact, the good average performance of the LSTM reported at horizon 3 is purely driven by the GFC episode, where the $AR(p)$ is particularly bad.

Finally, results from the FF models are mixed. For instance, during the 1993-2006 OOS period, the FF-cpi (and the autoregressive model) cannot track well the downward trend in inflation in the middle part of the sample at horizon 24, while the FF-pool is more successful. On the other hand, during 2006-2019, it is the FF-pool that shows the worst performance at horizon 24, where it underestimates inflation over a significant part of the sample. 

Overall, the recursive structure inherent to the LSTM, when compared to FF networks, appears to grant more robustness to predict at long horizons, but can hurt its performance at the medium term. Results on FF models are less conclusive, although their performance remains competitive against benchmarks. An important takeaway in terms of forecast performance is the sensitivity with respect to network architecture, and the practitioner solely interested in forecast performance would be advised to consider ensemble techniques that decrease the variance of single picked architectures, even if those are selected over a validation sample.

\section{Uncertainty over network initialization}
\label{sec:iv}

Neural network predictions can in general be sensitive to the the choice of initialization as well as network architecture as discussed above. In fact, machine learning models more generally are prone to instabilities in performance due to their sensitivity to model specification. \cite{damour2022} refers to this issue as underspecification and caution on the transferability of these models from the lab to real-world applications (see also \citealp{geirhos2020} and \citealp{buckmann2022} for similar discussions).

In this section, I contribute to this debate by focusing on the dependence of network predictions on initial values. As discussed previously, ensembles are usually appropriate in this case to smooth out predictions. Natural questions that emerge are how different are the performances of forecasts embedding different initializations, and more importantly how the performance of the ensemble forecast $\{ \hat{y}_{ens,t+h} \}_{t=1}^P$ compares with the performance of individual forecasts $\{ \hat{y}_{k,t+h} \}_{t=1}^P$, for $k=1,...,K$. To address these points, I compute a standard DM test statistic of equal forecast accuracy between the individual forecasts $k=1,...,K$ and the ensemble forecast. This exercise essentially yields a distribution of the DM test statistic over specifications embedding different initial values. 

Let $\{ e_{k,t+h} \}_{t=1}^P$ and $\{ e_{ens,t+h} \}_{t=1}^P$ be forecast errors associated with forecast $\{ \hat{y}_{k,t+h} \}_{t=1}^P$ and with the ensemble forecast $\{ \hat{y}_{ens,t+h} \}_{t=1}^P$ respectively over the OOS period. Now consider a loss-differential series $\{ d_{k,t+h} \}_{t=1}^P$, where $d_{k,t+h} \equiv [e_{k,t+h}^2 - e_{ens,t+h}^2]$, such that positive values are associated with a better performance of the ensemble prediction. For a given horizon $h$, I compute 
\begin{equation}
    \Delta_{k,h} = \frac{\Bar{d}_{k,h}}{\hat{\sigma}_d/\sqrt{P}}, \qquad k=1,...,K,
\label{eq:dm_stat}
\end{equation}
where $\Bar{d}_{k,h} = \frac{1}{P} \sum_{t=1}^P d_{k,t+h}$ is the sample mean loss-differential, and $\hat{\sigma_d}^2$ is a heteroskedasticity and autocorrelation consistent (HAC) estimate of the variance of $\Bar{d}_{k,h}$.

I discuss two potential issues with the use of the DM statistic on nestedness and parameter estimation error. For the former, the statistic seems appropriate as the two series being compared here can be better described as coming from different forecast strategies rather than models (see discussion in \citealp{diebold15}). Indeed, on the one hand we have a forecast from a neural net with initialization $k$ and on the other a forecast equal to the mean of $K$ different neural net forecasts, each with a different initialization. However, to alleviate the concern of nestedness regarding the use of the DM statistic, I also consider the mean squared error-adjusted test from \cite{clark_west07}, which is equivalent to an encompassing test and suitable in the context of nested models. This extended result is shown in Appendix \ref{app:iv_app} and briefly discussed below. On parameter estimation error, I note that the implied assumption from \cite{diebold} that the ratio $P/R$ vanishes asymptotically is violated, as here $P/R \approx 0.50$ or larger. However, as pointed out in \cite{corradi}, if the same loss function is used for both estimation and model selection purposes, which is the case in the paper (mean squared error loss), the contribution of parameter estimation error in the DM statistic becomes negligible.

\begin{figure}
    \centering
    \caption{Distribution of the DM statistic over different initializations}
    \includegraphics[width=16cm]{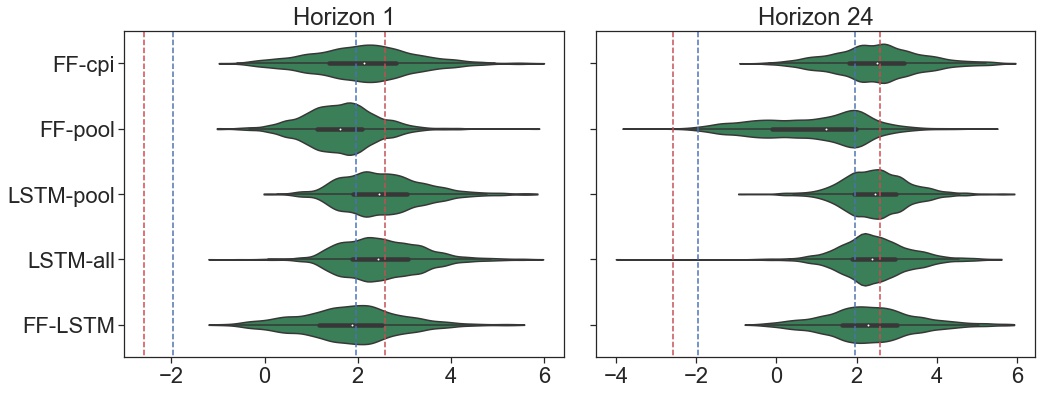}
    \caption*{\small DM test statistic $\Delta_{k,h}$, for $k=1,...,1400$ and horizons $h=1,24$. Positive values imply better forecast performance of the ensemble prediction over predictions from a particular initialization. Vertical dashed lines are Gaussian critical values at the $5\%$ (blue) and $1\%$ (red) significance levels. Values are truncated at 6 for visualization purposes.}
\label{fig:DM}
\end{figure}

I carry out the analysis considering the neural net models estimated in the previous section (as defined in Table \ref{table:hyper}), for $h=1$ and $h=24$, and $K=1400$ different initializations as before. Figure \ref{fig:DM} plots the distribution of the statistic $\Delta_{k,h}$ over $k=1,...,K$ for each model. Also plotted are the Gaussian critical values at the $5\%$ (blue dashed line) and $1\%$ (red dashed line) significance levels. 

The plots show that, in most cases, the means of the DM distributions lie beyond the $5\%$ line, indicating that the ensemble is on average better than any single initialization at the $5\%$ level of significance. In fact, on average across models and horizons, around $60\%$ of different initializations are statistically worse than the ensemble at the $5\%$ level, while the remaining $40\%$ (except for a few outliers) are not significantly different from the ensemble prediction. Additionally, there are no large differences across models. The same conclusions follow if one looks at the mean squared error-adjusted test, instead of the DM test, shown in Figure \ref{fig:DMadj} in the appendix. As expected, in this later case we observe higher values of the statistics, as the test corrects for the potential upward bias of the alternative model (here, the ensemble) under the null that its additional parameters are zero.

This exercise also reveals that predictions from models with different initial values can be substantially different (note the large variation of the test statistic across $k$'s).\footnote{\cite{buckmann2022} have similar findings, where the authors show that neural networks and support vector machines can be particularly sensitive to initial conditions compared to other machine learning models.} Individual predictions of the FF-pool, in particular, exhibit substantial variation (Figure \ref{fig:DM-other-hor} shows that the same is true for other horizons). Figure \ref{fig:lossdiff-dist} (in Appendix \ref{app:iv_app}) disentangles the contributions of the loss-differential mean and standard deviation within the DM statistic. Note that predictions from the FF-pool can be much more sensitive to the choice of initialization compared to the FF-cpi. Both the mean and standard deviation of the loss-differential can take on relatively high values, which explains the higher dispersion in DM statistics of the FF-pool compared to other models. Interestingly, the salient difference between the FF-pool and the FF-cpi is the number of parameters (Table \ref{table:hyper}), suggesting that larger networks, all else equal, can create additional instabilities in model predictions.

Overall, this evidence substantiates the use of neural network ensembles given the sensitivity to initial values and network architecture (Section \ref{sec:OFS} also discusses the dependence of forecast performance on architecture). Additionally, due to the highly nonlinear setting, it is plausible that the best initial values change over time. Combining the outcome of different forecasts in an ensemble-like approach is therefore a way of making the final prediction more robust to this type of uncertainty.

\section{Concluding remarks}
\label{sec:conclusions}

This paper investigates the suitability of a recurrent neural network, specifically the LSTM, to forecast inflation, while analysing advantages and drawbacks of using this model in terms of interpretability and implementation. On forecast performance, LSTM-based models are good at forecasting the long-term trend of the inflation series and perform well during periods of higher macroeconomic uncertainty, such as during the post-GFC period. This is due to their lower sensitivity to temporary and sudden price changes compared to traditional models in the literature. However, their overall forecasting performance is not outstanding, especially compared to the random forest model. The analysis also reveals that factors generated by the estimation of LSTM models capture the long-term trend of inflation, along with output gap-type signals, which is consistent with a Phillips curve-based specification. Finally, the paper sheds light on the sensitivity of neural network's predictions to initial values and network size. Specifically, networks of different sizes generally result in different predictions, and larger networks do not necessarily outperform smaller ones. Additionally, neural networks' predictions are significantly sensitive to the initialization, where large enough networks can exacerbate this sensitivity even more. Practitioners interested in forecasting performance should, therefore, consider the use of ensemble techniques to average out the variation from both initial values and network size (or network architecture in general), which significantly stabilizes predictions.

\bibliography{bib}

\begin{thebibliography}{56}
\def\enquote#1{``#1''}
\expandafter\ifx\csname natexlab\endcsname\relax\def\natexlab#1{#1}\fi

\bibitem[{Almosova and Andresen(2019)}]{almosova}
Almosova, A., and N.~Andresen.  (2019). \enquote{Nonlinear inflation
  forecasting with recurrent neural networks.} .

\bibitem[{Andreini et~al.(2020)Andreini, Izzo, and Ricco}]{andreini}
Andreini, P., C.~Izzo, and G.~Ricco.  (2020). \enquote{Deep dynamic factor
  models.} arXiv:2007.11887 .

\bibitem[{Athey(2019)}]{athey}
Athey, S.  (2019). \enquote{The Impact of Machine Learning on Economics.} In
  \enquote{Ajay Agrawal, Joshua Gans, and Avi Goldfarb (Eds.), The Economics of
  Artificial Intelligence: An Agenda,} 507--547, Chicago: University of Chicago
  Press.

\bibitem[{Atkeson and Ohanian(2001)}]{ao}
Atkeson, A., and L.~E. Ohanian.  (2001). \enquote{Are Phillips curves useful
  for forecasting inflation?} Quarterly Review, Federal Reserve Bank of
  Minneapolis 2--11.

\bibitem[{Bai and Ng(2008)}]{bai_ng}
Bai, J., and S.~Ng.  (2008). \enquote{Forecasting economic time series using
  targeted predictors.} Journal of Econometrics 146, 2, 304--317.

\bibitem[{Breiman(2001)}]{breiman2001}
Breiman, L..  (2001). \enquote{Random Forests.} Machine Learning 45, 5--32.

\bibitem[{Breiman et~al.(1984)Breiman, Friedman, Olshen, and Stone}]{breiman84}
Breiman, L., J.~H. Friedman, R.~A. Olshen, and C.~G. Stone (1984).
  Classification and regression trees. Wadsworth International Group, Belmont,
  California, USA.

\bibitem[{Buckmann and Joseph(2022)}]{buckmann2022}
Buckmann, M., and A.~Joseph.  (2022). \enquote{An interpretable machine
  learning workflow with an application to economic forecasting.} Bank of
  England Working Paper No 984 .

\bibitem[{Chakraborty and Joseph(2017)}]{joseph}
Chakraborty, C., and A.~Joseph.  (2017). \enquote{Machine learning at central
  banks.} Bank of England Working Papers 674.

\bibitem[{Choudhary and Haider(2012)}]{choudhary}
Choudhary, M.~S., and A.~Haider.  (2012). \enquote{Neural network models for
  inflation forecasting: an appraisal.} Applied Economics 44, 20, 2631--2635.

\bibitem[{Clark and West(2007)}]{clark_west07}
Clark, T.~E., and K.~D. West.  (2007). \enquote{Approximately normal tests for
  equal predictive accuracy in nested models.} Journal of Econometrics 138,
  291--311.

\bibitem[{Cook and Hall(2017)}]{cook}
Cook, T.~R., and S.~Hall.  (2017). \enquote{Macroeconomic Indicator Forecasting
  with Deep Neural Networks.} In \enquote{Federal Reserve Bank of Kansas City,
  Research Working Paper 17-11,} .

\bibitem[{Corradi and Swanson(2006)}]{corradi}
Corradi, V., and N.~R. Swanson.  (2006). \enquote{Chapter 5 Predictive Density
  Evaluation.} volume~1 of Handbook of Economic Forecasting, 197--284,
  Elsevier.

\bibitem[{Coulombe(2022)}]{coulombePC}
Coulombe, P.~G..  (2022). \enquote{A neural Phillips curve and a deep output
  gap.} Available at SSRN: https://ssrn.com/abstract=4018079 .

\bibitem[{Coulombe et~al.(2020)Coulombe, Leroux, Stevanovic, and
  Surprenant}]{coulombe2020machine}
Coulombe, P.~G., M.~Leroux, D.~Stevanovic, and S.~Surprenant.  (2020).
  \enquote{How is Machine Learning Useful for Macroeconomic Forecasting?}
  arXiv:2008.12477 .

\bibitem[{D'Amour et~al.(2022)D'Amour, Heller, Moldovan et~al.}]{damour2022}
D'Amour, A., K.~Heller, D.~Moldovan, et~al..  (2022).
  \enquote{Underspecification presents challenges for credibility in modern
  machine learning.} Journal of Machine Learning Research 23, 1--61.

\bibitem[{Diebold and Mariano(1995)}]{diebold}
Diebold, F., and R.~Mariano.  (1995). \enquote{Comparing Predictive Accuracy.}
  Journal of Business \& Economic Statistics 13, 3, 253--263.

\bibitem[{Diebold(2015)}]{diebold15}
Diebold, F.~X..  (2015). \enquote{Comparing Predictive Accuracy, Twenty Years
  Later: A Personal Perspective on the Use and Abuse of Diebold-Mariano Tests.}
  Journal of Business \& Economic Statistics 33, 1, 1--1.

\bibitem[{Exterkate et~al.(2016)Exterkate, Groenen, Heij, and
  Van~Dijk}]{EXTERKATE}
Exterkate, P., P.J.F. Groenen, C.~Heij, and D.~Van~Dijk.  (2016).
  \enquote{Nonlinear forecasting with many predictors using kernel ridge
  regression.} International Journal of Forecasting 32, 3, 736--753.

\bibitem[{Farrell et~al.(2021)Farrell, Liang, and Misra}]{farrell}
Farrell, M.~H., T.~Liang, and S.~Misra.  (2021). \enquote{Deep neural networks
  for estimation and inference.} Econometrica 89, 1, 181--213.

\bibitem[{Faust and Wright(2013)}]{faust}
Faust, J., and J.~Wright.  (2013). \enquote{Forecasting Inflation.} In
  \enquote{Handbook of Economic Forecasting,} volume~2A, Elsevier.

\bibitem[{Fort et~al.(2020)Fort, Hu, and Lakshminarayanan}]{fort2020}
Fort, S., H.~Hu, and B.~Lakshminarayanan.  (2020). \enquote{Deep Ensembles: A
  Loss Landscape Perspective.}

\bibitem[{Geirhos et~al.(2020)Geirhos, Jacobsen, Michaelis
  et~al.}]{geirhos2020}
Geirhos, R., J-H. Jacobsen, C.~Michaelis, et~al..  (2020). \enquote{Shortcut
  learning in deep neural networks.} Nature Machine Intelligence 2, 665--673.

\bibitem[{Giacomini and Rossi(2010)}]{rossi}
Giacomini, R., and B.~Rossi.  (2010). \enquote{Forecast Comparisons in Unstable
  Environments.} Journal of Applied Econometrics 25, 595--620.

\bibitem[{Giannone et~al.(2015)Giannone, Lenza, and Primiceri}]{giannone}
Giannone, D., M.~Lenza, and G.~Primiceri.  (2015). \enquote{Prior selection for
  vector autoregressions.} Review of Economics and Statistics 97, 2, 436--451.

\bibitem[{Glorot and Bengio(2010)}]{glorot}
Glorot, X., and Y.~Bengio.  (2010). \enquote{Understanding the difficulty of
  training deep feedforward neural networks.} Journal of Machine Learning
  Research - Proceedings Track 9, 249--256.

\bibitem[{Goodfellow et~al.(2016)Goodfellow, Bengio, and
  Courville}]{Goodfellow-et-al-2016}
Goodfellow, I.~J., Y.~Bengio, and A.~Courville (2016). Deep Learning. MIT
  Press, \url{http://www.deeplearningbook.org}.

\bibitem[{Gu et~al.(2019)Gu, Kelly, and Xiu}]{gu}
Gu, S., B.~Kelly, and D.~Xiu.  (2019). \enquote{Empirical Asset Pricing via
  Machine Learning.} In \enquote{Chicago Booth Research Paper No. 18-04; 31st
  Australasian Finance and Banking Conference 2018; Yale ICF Working Paper No.
  2018-09.}, .

\bibitem[{Hanin and Rolnick(2018)}]{hanin}
Hanin, B., and D.~Rolnick.  (2018). \enquote{How to start training: The effect
  of initialization and architecture.} In \enquote{Advances in Neural
  Information Processing Systems 31,} 569--579, Curran Associates, Inc.

\bibitem[{Hansen et~al.(2011)Hansen, Lunde, and Nason}]{hansen_mcs}
Hansen, P.~R., A.~Lunde, and J.~M. Nason.  (2011). \enquote{The model
  confidence set.} Econometrica 79, 2, 453--497.

\bibitem[{Hasenzagl et~al.(2022)Hasenzagl, Pellegrino, Reichlin, and
  Ricco}]{gio}
Hasenzagl, T., F.~Pellegrino, L.~Reichlin, and G.~Ricco.  (2022). \enquote{A
  Model of the Fed's View on Inflation.} The Review of Economics and Statistics
  104, 4.

\bibitem[{Hastie et~al.(2001)Hastie, Tibshirani, and
  Friedman}]{hastie01statisticallearning}
Hastie, T., R.~Tibshirani, and J.~Friedman (2001). The Elements of Statistical
  Learning. Springer Series in Statistics, New York, NY, USA: Springer New York
  Inc.

\bibitem[{Hauzenberger et~al.(2023)Hauzenberger, Huber, and
  Klieber}]{hauzenberger}
Hauzenberger, N., F.~Huber, and K.~Klieber.  (2023). \enquote{Real-time
  inflation forecasting using non-linear dimension reduction techniques.}
  International Journal of Forecasting 39, 2, 901--921.

\bibitem[{He et~al.(2015)He, Zhang, S., and Sun}]{he}
He, K., X.~Zhang, Ren S., and J.~Sun.  (2015). \enquote{Delving deep into
  rectifiers: surpassing human-level performance on imagenet classification.}
  In \enquote{Proceedings of the IEEE international conference on computer
  vision.}, 1026--1034.

\bibitem[{Hinton and Salakhutdinov(2006)}]{hinton}
Hinton, G.~E., and R.~R. Salakhutdinov.  (2006). \enquote{Reducing the
  Dimensionality of Data with Neural Networks.} Science 313, 5786, 504--507.

\bibitem[{Ioffe and Szegedy(2015)}]{norm}
Ioffe, S., and C.~Szegedy.  (2015). \enquote{Batch Normalization: Accelerating
  Deep Network Training by Reducing Internal Covariate Shift.} arXiv:1502.03167
  .

\bibitem[{Izmailov et~al.(2019)Izmailov, Podoprikhin, Garipov, Vetrov, and
  Wilson}]{izmailov2019}
Izmailov, P., D.~Podoprikhin, T.~Garipov, D.~Vetrov, and A.~G. Wilson.  (2019).
  \enquote{Averaging Weights Leads to Wider Optima and Better Generalization.}

\bibitem[{Jain and Kar(2017)}]{jain}
Jain, P., and P.~Kar.  (2017). \enquote{Non-convex Optimization for Machine
  Learning.} Foundations and Trends in Machine Learning 10, 3-4, 142--336.

\bibitem[{Kingma and Ba(2015)}]{adam}
Kingma, D., and J.~Ba.  (2015). \enquote{Adam: A method for stochastic
  optimization.} In \enquote{ICLR,} .

\bibitem[{Kuan and White(1994)}]{kuan}
Kuan, C-M., and H.~White.  (1994). \enquote{Artificial neural networks: an
  econometric perspective.} Econometric Reviews 13, 1, 1--91.

\bibitem[{Ludvigson and Ng(2007)}]{lud}
Ludvigson, S., and S.~Ng.  (2007). \enquote{The empirical risk return relation:
  A factor analysis approach.} Journal of Financial Economics 83, 1, 171--222.

\bibitem[{McAdam and McNelis(2005)}]{MCADAM}
McAdam, P., and P.~McNelis.  (2005). \enquote{Forecasting inflation with thick
  models and neural networks.} Economic Modelling 22, 5, 848--867.

\bibitem[{McCracken and Ng(2016)}]{fred}
McCracken, M.~W., and S.~Ng.  (2016). \enquote{FRED-MD: A monthly database for
  Macroeconomic Research.} Journal of Business \& Economic Statistics 34, 4,
  574--589.

\bibitem[{Medeiros et~al.(2019)Medeiros, Vasconcelos, Veiga, and
  Zilberman}]{medeiros}
Medeiros, M.~C., G.~Vasconcelos, A.~Veiga, and E.~Zilberman.  (2019).
  \enquote{Forecasting Inflation in a Data-Rich Environment: The Benefits of
  Machine Learning Methods.} Journal of Business \& Economic Statistics 1--45.

\bibitem[{Mullainathan and Spiess(2017)}]{mullainathan}
Mullainathan, S., and J.~Spiess.  (2017). \enquote{Machine Learning: An Applied
  Econometric Approach.} Journal of Economic Perspectives 31, 2, 87--106.

\bibitem[{Nakamura(2005)}]{nakamura}
Nakamura, E.  (2005). \enquote{Inflation forecasting using a neural network.}
  Economics Letters 86, 373--378.

\bibitem[{Refenes and White(1998)}]{refenes&white}
Refenes, A.P., and H.~White.  (1998). \enquote{Neural Networks and Financial
  Economics.} International Journal of Forecasting 17, 347–495.

\bibitem[{Schmidt-Hieber(2020)}]{schmidt}
Schmidt-Hieber, J.  (2020). \enquote{Nonparametric regression using deep neural
  networks with ReLU activation function.} The Annals of Statistics 48, 4,
  1875--1897.

\bibitem[{Sermpinis et~al.(2014)Sermpinis, Stasinakis, Theofilatos, and
  Karathanasopoulos}]{sermpinis}
Sermpinis, G., C.~Stasinakis, K.~Theofilatos, and A.~Karathanasopoulos.
  (2014). \enquote{Inflation and Unemployment Forecasting with Genetic Support
  Vector Regression.} Journal of Forecasting 33, 471–487.

\bibitem[{Stock and Watson(1999)}]{stock99}
Stock, J.~H., and M.~W. Watson.  (1999). \enquote{Forecasting Inflation.}
  Journal of Monetary Economics 44, 293--335.

\bibitem[{Stock and Watson(2002)}]{stock02}
Stock, J.~H., and M.~W. Watson.  (2002). \enquote{Macroeconomic forecasting
  with diffusion indexes.} Journal of Business \& Economic Statistics 20,
  147--162.

\bibitem[{Stock and Watson(2007)}]{stock07}
Stock, J.~H., and M.~W. Watson.  (2007). \enquote{Why Has U.S. Inflation Become
  Harder to Forecast?} Journal of Money, Credit and Banking. 39, 7, 1849--1849.

\bibitem[{Stock and Watson(2009)}]{stock09}
Stock, J.~H., and M.~W. Watson.  (2009). \enquote{Phillips Curve Inflation
  Forecasts.} In \enquote{Fuhrer J, Kodrzycki Y, Little J, Olivei G
  Understanding Inflation and the Implications for Monetary Policy.}, 99--202,
  Cambridge: MIT Press.

\bibitem[{Varian(2014)}]{varian}
Varian, H.~R.  (2014). \enquote{Big data: New tricks for econometrics.} The
  Journal of Economic Perspectives 28, 2, 3--27.

\bibitem[{Verstyuk(2020)}]{verstyuk}
Verstyuk, S..  (2020). \enquote{Modeling multivariate time series in economics:
  From auto-regressions to recurrent neural networks.} Available at SSRN
  3589337 .

\bibitem[{Wilson and Izmailov(2022)}]{wilson2022}
Wilson, A.~G., and P.~Izmailov.  (2022). \enquote{Bayesian Deep Learning and a
  Probabilistic Perspective of Generalization.}

\end{thebibliography}
\bibliographystyle{mystyle} 

\appendix
\clearpage
\begin{center}
\textbf{\Large Supplementary Materials: \\ Predicting Inflation with Recurrent Neural Networks}
\end{center}
\setcounter{equation}{0}
\setcounter{figure}{0}
\setcounter{table}{0}
\setcounter{page}{1}
\setcounter{section}{0}
\makeatletter
\renewcommand{\theequation}{S\arabic{equation}}
\renewcommand{\thefigure}{S\arabic{figure}}
\renewcommand{\thetable}{S\arabic{table}}
\renewcommand{\bibnumfmt}[1]{[S#1]}
\renewcommand{\citenumfont}[1]{S#1}

\vspace{-2cm}

\addcontentsline{toc}{section}{Appendix} 
\part{} 
\parttoc 

\begin{landscape}
\section{Data description} \label{app:data}
\begingroup
\footnotesize
\renewcommand\arraystretch{1}

\begin{ThreePartTable}
\renewcommand\TPTminimum{\textwidth}
\setlength\LTleft{10pt}
\setlength\LTright{10pt}
\setlength\tabcolsep{8pt}

\begin{TableNotes} \small \setstretch{1.1}
\item The table describes the data used in the empirical analysis, collected from the FRED monthly database using the October-2019 vintage. I follow \cite{fred} and divide the series into eight economic groups. The column Tcode refers to the transformation applied to each series $x_t$, where (1) no transformation, (2) $\Delta x_t$, (3) $\Delta^2 x_t$, (4) $log(x_t)$, (5) $\Delta log(x_t)$, (6) $\Delta^2 log(x_t)$, (7) $\Delta (x_t/x_{t-1} - 1.0)$. The comparable series in Global Insight is given in the column GSI. Percentage of missing values with respect to the full sample (Jan-59 to Oct-19) is included in the last column.
\vspace{5mm}
\end{TableNotes}

\begin{longtable}[c]{ccclclc}
  \caption{Data description}  \\
  \insertTableNotes \\
  & Tcode & \multicolumn{1}{c}{Fred mnemonics} & \multicolumn{1}{c}{Description} & GSI   & \multicolumn{1}{c}{GSI: description} & \multicolumn{1}{c}{Missing ($\%$)} \\
 \hline
  \endfirsthead
  \multicolumn{6}{c}%
  {\tablename\ \thetable\ -- \textit{Continued}} \\
 & Tcode & \multicolumn{1}{c}{Fred mnemonics} & \multicolumn{1}{c}{Description} & GSI   & \multicolumn{1}{c}{GSI: description} & \multicolumn{1}{c}{Missing ($\%$)} \\ \hline
  \endhead
  \hline \multicolumn{7}{r}{\textit{Continued on next page}} \\
  \endfoot
  \hline
  \endlastfoot
    \multicolumn{6}{l}{\textit{Group 1: Output and income}} & \\
    1     & 5     & RPI   & Real Personal Income & M\_14386177 & \multicolumn{1}{l}{PI} & \\
    2     & 5     & W875RX1 & Real personal income ex transfer receipts & M\_145256755 & \multicolumn{1}{l}{PI less transfers} & \\
    3     & 5     & INDPRO & IP Index & M\_116460980 & \multicolumn{1}{l}{IP: total} & \\
    4     & 5     & IPFPNSS & IP: Final Products and Nonindustrial Supplies & M\_116460981 & \multicolumn{1}{l}{IP: products} & \\
    5     & 5     & IPFINAL & IP: Final Products (Market Group) & M\_116461268 & \multicolumn{1}{l}{IP: final prod} & \\
    6     & 5     & IPCONGD & IP: Consumer Goods & M\_116460982 & \multicolumn{1}{l}{IP: cons gds} & \\
    7     & 5     & IPDCONGD & IP: Durable Consumer Goods & M\_116460983 & \multicolumn{1}{l}{IP: cons dble} & \\
    8     & 5     & IPNCONGD & IP: Nondurable Consumer Goods & M\_116460988 & \multicolumn{1}{l}{IP: cons nondble} & \\
    9     & 5     & IPBUSEQ & IP: Business Equipment & M\_116460995 & \multicolumn{1}{l}{IP: bus eqpt} & \\
    10    & 5     & IPMAT & IP: Materials & M\_116461002 & \multicolumn{1}{l}{IP: matls} & \\
    11    & 5     & IPDMAT & IP: Durable Materials & M\_116461004 & \multicolumn{1}{l}{IP: dble matls} & \\
    12    & 5     & IPNMAT & IP: Nondurable Materials & M\_116461008 & \multicolumn{1}{l}{IP: nondble matls} & \\
    13    & 5     & IPMANSICS & IP: Manufacturing (SIC) & M\_116461013 & \multicolumn{1}{l}{IP: mfg} & \\
    14    & 5     & IPB51222s & IP: Residential Utilities & M\_116461276 & \multicolumn{1}{l}{IP: res util} & \\
    15    & 5     & IPFUELS & IP: Fuels & M\_116461275 & \multicolumn{1}{l}{IP: fuels} & \\
    16    & 2     & CUMFNS & Capacity Utilization: Manufacturing & M\_116461602 & \multicolumn{1}{l}{Cap util} & \\
    \multicolumn{6}{l}{\textit{Group 2: Labor market}}  &  \\
    17    & 2     & HWI   & Help-Wanted Index for United States &       & \multicolumn{1}{l}{Help wanted indx } & 0.14 \\
    18    & 2     & HWIURATIO & Ratio of Help Wanted/No. Unemployed & M\_110156531 & \multicolumn{1}{l}{Help wanted/unemp} &  0.14 \\
    19    & 5     & CLF16OV & Civilian Labor Force & M\_110156467 & \multicolumn{1}{l}{Emp CPS total} & \\
    20    & 5     & CE16OV & Civilian Employment & M\_110156498 & \multicolumn{1}{l}{Emp CPS nonag} & \\
    21    & 2     & UNRATE & Civilian Unemployment Rate & M\_110156541 & \multicolumn{1}{l}{U: all} & \\
    22    & 2     & UEMPMEAN & Average Duration of Unemployment (Weeks) & M\_110156528 & \multicolumn{1}{l}{U: mean duration} & \\
    23    & 5     & UEMPLT5 & Civilians Unemployed - Less Than 5 Weeks & M\_110156527 & \multicolumn{1}{l}{U < 5 wks} & \\
    24    & 5     & UEMP5TO14 & Civilians Unemployed for 41760 Weeks & M\_110156523 & \multicolumn{1}{l}{U 41760 wks} & \\
    25    & 5     & UEMP15OV & Civilians Unemployed - 15 Weeks \& Over & M\_110156524 & \multicolumn{1}{l}{U 15+ wks} & \\
    26    & 5     & UEMP15T26 & Civilians Unemployed for 15-26 Weeks & M\_110156525 & \multicolumn{1}{l}{U 15-26 wks} & \\
    27    & 5     & UEMP27OV & Civilians Unemployed for 27 Weeks and Over & M\_110156526 & \multicolumn{1}{l}{U 27+ wks} & \\
    28    & 5     & CLAIMSx & Initial Claims & M\_15186204 & \multicolumn{1}{l}{UI claims} & \\
    29    & 5     & PAYEMS & All Employees: Total nonfarm & M\_123109146 & \multicolumn{1}{l}{Emp: total}&  \\
    30    & 5     & USGOOD & All Employees: Goods-Producing Industries & M\_123109172 & \multicolumn{1}{l}{Emp: gds prod} & \\
    31    & 5     & CES1021000001 & All Employees: Mining and Logging: Mining & M\_123109244 & \multicolumn{1}{l}{Emp: mining}&  \\
    32    & 5     & USCONS & All Employees: Construction & M\_123109331 & \multicolumn{1}{l}{Emp: const} & \\
    33    & 5     & MANEMP & All Employees: Manufacturing & M\_123109542 & \multicolumn{1}{l}{Emp: mfg} & \\
    34    & 5     & DMANEMP & All Employees: Durable goods & M\_123109573 & \multicolumn{1}{l}{Emp: dble gds} & \\
    35    & 5     & NDMANEMP & All Employees: Nondurable goods & M\_123110741 & \multicolumn{1}{l}{Emp: nondbles} & \\
    36    & 5     & SRVPRD & All Employees: Service-Providing Industries & M\_123109193 & \multicolumn{1}{l}{Emp: services} & \\
    37    & 5     & USTPU & All Employees: Trade, Transportation \& Utilities & M\_123111543 & \multicolumn{1}{l}{Emp: TTU} & \\
    38    & 5     & USWTRADE & All Employees: Wholesale Trade & M\_123111563 & \multicolumn{1}{l}{Emp: wholesale} & \\
    39    & 5     & USTRADE & All Employees: Retail Trade & M\_123111867 & \multicolumn{1}{l}{Emp: retail} & \\
    40    & 5     & USFIRE & All Employees: Financial Activities & M\_123112777 & \multicolumn{1}{l}{Emp: FIRE} & \\
    41    & 5     & USGOVT & All Employees: Government & M\_123114411 & \multicolumn{1}{l}{Emp: Govt} & \\
    42    & 1     & CES0600000007 & Avg Weekly Hours : Goods-Producing & M\_140687274 & \multicolumn{1}{l}{Avg hrs} & \\
    43    & 2     & AWOTMAN & Avg Weekly Overtime Hours : Manufacturing & M\_123109554 & \multicolumn{1}{l}{Overtime: mfg} & \\
    44    & 1     & AWHMAN & Avg Weekly Hours : Manufacturing & M\_14386098 & \multicolumn{1}{l}{Avg hrs: mfg} & \\
    45    & 6     & CES0600000008 & Avg Hourly Earnings : Goods-Producing & M\_123109182 & \multicolumn{1}{l}{AHE: goods} & \\
    46    & 6     & CES2000000008 & Avg Hourly Earnings : Construction & M\_123109341 & \multicolumn{1}{l}{AHE: const} & \\
    47    & 6     & CES3000000008 & Avg Hourly Earnings : Manufacturing & M\_123109552 & \multicolumn{1}{l}{AHE: mfg} & \\
    \multicolumn{6}{l}{\textit{Group 3: Housing}}  & \\
    48    & 4     & HOUST & Housing Starts: Total New Privately Owned & M\_110155536 & \multicolumn{1}{l}{Starts: nonfarm} & \\
    49    & 4     & HOUSTNE & Housing Starts, Northeast & M\_110155538 & \multicolumn{1}{l}{Starts: NE} & \\
    50    & 4     & HOUSTMW & Housing Starts, Midwest & M\_110155537 & \multicolumn{1}{l}{Starts: MW}&  \\
    51    & 4     & HOUSTS & Housing Starts, South & M\_110155543 & \multicolumn{1}{l}{Starts: South} & \\
    52    & 4     & HOUSTW & Housing Starts, West & M\_110155544 & \multicolumn{1}{l}{Starts: West} & \\
    53    & 4     & PERMIT & New Private Housing Permits (SAAR) & M\_110155532 & \multicolumn{1}{l}{BP: total} & 1.64\\
    54    & 4     & PERMITNE & New Private Housing Permits, Northeast (SAAR) & M\_110155531 & \multicolumn{1}{l}{BP: NE} & 1.64\\
    55    & 4     & PERMITMW & New Private Housing Permits, Midwest (SAAR) & M\_110155530 & \multicolumn{1}{l}{BP: MW} & 1.64\\
    56    & 4     & PERMITS & New Private Housing Permits, South (SAAR) & M\_110155533 & \multicolumn{1}{l}{BP: South} & 1.64\\
    57    & 4     & PERMITW & New Private Housing Permits, West (SAAR) & M\_110155534 & \multicolumn{1}{l}{BP: West} & 1.64\\
    \multicolumn{6}{l}{\textit{Group 4: Consumption, orders and inventories}}   & \\
    58    & 5     & DPCERA3M086SBEA & Real personal consumption expenditures & M\_123008274 & \multicolumn{1}{l}{Real Consumption} & \\
    59    & 5     & CMRMTSPLx & Real Manu. and Trade Industries Sales & M\_110156998 & \multicolumn{1}{l}{M\&T sales} & 0.14 \\
    60    & 5     & RETAILx & Retail and Food Services Sales & M\_130439509 & \multicolumn{1}{l}{Retail sales} & \\
    61    & 5     & ACOGNO & New Orders for Consumer Goods & M\_14385863 & \multicolumn{1}{l}{Orders: cons gds} & 54.52 \\
    62    & 5     & AMDMNOx & New Orders for Durable Goods & M\_14386110 & \multicolumn{1}{l}{Orders: dble gds} & \\
    63    & 5     & ANDENOx & New Orders for Nondefense Capital Goods & M\_178554409 & \multicolumn{1}{l}{Orders: cap gds} & 14.93 \\
    64    & 5     & AMDMUOx & Unfilled Orders for Durable Goods & M\_14385946 & \multicolumn{1}{l}{Unf orders: dble} & \\
    65    & 5     & BUSINVx & Total Business Inventories & M\_15192014 & \multicolumn{1}{l}{M\&T invent}&  0.14 \\
    66    & 2     & ISRATIOx & Total Business: Inventories to Sales Ratio & M\_15191529 & \multicolumn{1}{l}{M\&T invent/sales} & 0.14 \\
    67    & 2     & UMCSENTx & Consumer Sentiment Index & hhsntn & \multicolumn{1}{l}{Consumer expect} & 21.10 \\
    \multicolumn{6}{l}{\textit{Group 5:  Money and credit}}  & \\
    68    & 6     & M1SL  & M1 Money Stock & M\_110154984 & \multicolumn{1}{l}{M1} & \\
    69    & 6     & M2SL  & M2 Money Stock & M\_110154985 & \multicolumn{1}{l}{M2} & \\
    70    & 5     & M2REAL & Real M2 Money Stock & M\_110154985 & \multicolumn{1}{l}{M2 (real)} & \\
    71    & 6     & BOGMBASE & Monetary Base & M\_110154995 & \multicolumn{1}{l}{MB} & \\
    72    & 6     & TOTRESNS & Total Reserves of Depository Institutions & M\_110155011 & \multicolumn{1}{l}{Reserves tot} & \\
    73    & 7     & NONBORRES & Reserves Of Depository Institutions & M\_110155009 & \multicolumn{1}{l}{Reserves nonbor} & \\
    74    & 6     & BUSLOANS & Commercial and Industrial Loans & BUSLOANS & \multicolumn{1}{l}{C\&I loan plus} & \\
    75    & 6     & REALLN & Real Estate Loans at All Commercial Banks & BUSLOANS & \multicolumn{1}{l}{DC\&I loans} & \\
    76    & 6     & NONREVSL & Total Nonrevolving Credit & M\_110154564 & \multicolumn{1}{l}{Cons credit} & 0.14 \\
    77    & 2     & CONSPI & Nonrevolving consumer credit to Personal Income & M\_110154569 & \multicolumn{1}{l}{Inst cred/PI} & 0.14\\
    78    & 6     & MZMSL & MZM Money Stock & N.A.  & \multicolumn{1}{l}{N.A.} & \\
    79    & 6     & DTCOLNVHFNM & Consumer Motor Vehicle Loans Outstanding & N.A.  & \multicolumn{1}{l}{N.A.} & 0.14 \\
    80    & 6     & DTCTHFNM & Total Consumer Loans and Leases Outstanding & N.A.  & \multicolumn{1}{l}{N.A.} & 0.14 \\
    81    & 6     & INVEST & Securities in Bank Credit at All Commercial Banks & N.A.  & \multicolumn{1}{l}{N.A.} & \\
    \multicolumn{6}{l}{\textit{Group 6: Interest and exchange rates}} & \\
    82    & 2     & FEDFUNDS & Effective Federal Funds Rate & M\_110155157 & \multicolumn{1}{l}{Fed Funds} & \\
    83    & 2     & CP3Mx & 3-Month AA Financial Commercial Paper Rate & CPF3M & \multicolumn{1}{l}{Comm paper} & \\
    84    & 2     & TB3MS & 3-Month Treasury Bill: & M\_110155165 & \multicolumn{1}{l}{3 T-bill} & \\
    85    & 2     & TB6MS & 6-Month Treasury Bill: & M\_110155166 & \multicolumn{1}{l}{6 T-bill} & \\
    86    & 2     & GS1   & 1-Year Treasury Rate & M\_110155168 & \multicolumn{1}{l}{1 T-bond} & \\
    87    & 2     & GS5   & 5-Year Treasury Rate & M\_110155174 & \multicolumn{1}{l}{5 T-bond} & \\
    88    & 2     & GS10  & 10-Year Treasury Rate & M\_110155169 & \multicolumn{1}{l}{10 T-bond} & \\
    89    & 2     & AAA   & Moody’s Seasoned Aaa Corporate Bond Yield &       & \multicolumn{1}{l}{Aaa bond} & \\
    90    & 2     & BAA   & Moody’s Seasoned Baa Corporate Bond Yield &       & \multicolumn{1}{l}{Baa bond} & \\
    91    & 1     & COMPAPFFx & 3-Month Commercial Paper Minus FEDFUNDS &       & \multicolumn{1}{l}{CP-FF spread}&  \\
    92    & 1     & TB3SMFFM & 3-Month Treasury C Minus FEDFUNDS &       & \multicolumn{1}{l}{3 mo-FF spread} & \\
    93    & 1     & TB6SMFFM & 6-Month Treasury C Minus FEDFUNDS &       & \multicolumn{1}{l}{6 mo-FF spread} & \\
    94    & 1     & T1YFFM & 1-Year Treasury C Minus FEDFUNDS &       & \multicolumn{1}{l}{1 yr-FF spread} & \\
    95    & 1     & T5YFFM & 5-Year Treasury C Minus FEDFUNDS &       & \multicolumn{1}{l}{5 yr-FF spread} & \\
    96    & 1     & T10YFFM & 10-Year Treasury C Minus FEDFUNDS &       & \multicolumn{1}{l}{10 yr-FF spread} & \\
    97    & 1     & AAAFFM & Moody’s Aaa Corporate Bond Minus FEDFUNDS &       & \multicolumn{1}{l}{Aaa-FF spread} & \\
    98    & 1     & BAAFFM & Moody’s Baa Corporate Bond Minus FEDFUNDS &       & \multicolumn{1}{l}{Baa-FF spread} & \\
    99    & 5     & TWEXAFEGSMTHx & Trade Weighted U.S. Dollar Index &       & \multicolumn{1}{l}{Ex rate: avg} & 23.01 \\
    100   & 5     & EXSZUSx & Switzerland / U.S. Foreign Exchange Rate & M\_110154768 & \multicolumn{1}{l}{Ex rate: Switz} & \\
    101   & 5     & EXJPUSx & Japan / U.S. Foreign Exchange Rate & M\_110154755 & \multicolumn{1}{l}{Ex rate: Japan}&  \\
    102   & 5     & EXUSUKx & U.S. / U.K. Foreign Exchange Rate & M\_110154772 & \multicolumn{1}{l}{Ex rate: UK} & \\
    103   & 5     & EXCAUSx & Canada / U.S. Foreign Exchange Rate & M\_110154744 & \multicolumn{1}{l}{EX rate: Canada} & \\
    \multicolumn{6}{l}{\textit{Group 7: Prices}}  & \\
    104   & 6     & WPSFD49207 & PPI: Finished Goods & M110157517 & \multicolumn{1}{l}{PPI: fin gds} & \\
    105   & 6     & WPSFD49502 & PPI: Finished Consumer Goods & M110157508 & \multicolumn{1}{l}{PPI: cons gds} & \\
    106   & 6     & WPSID61 & PPI: Intermediate Materials & M\_110157527 & \multicolumn{1}{l}{PPI: int matls}&  \\
    107   & 6     & WPSID62 & PPI: Crude Materials & M\_110157500 & \multicolumn{1}{l}{PPI: crude matls} & \\
    108   & 6     & OILPRICEx & Crude Oil, spliced WTI and Cushing & M\_110157273 & \multicolumn{1}{l}{Spot market price} & \\
    109   & 6     & PPICMM & PPI: Metals and metal products: & M\_110157335 & \multicolumn{1}{l}{PPI: nonferrous}&  \\
    110   & 5     & CPIAUCSL & CPI : All Items & M\_110157323 & \multicolumn{1}{l}{CPI-U: all}&  \\
    111   & 6     & CPIAPPSL & CPI : Apparel & M\_110157299 & \multicolumn{1}{l}{CPI-U: apparel} & \\
    112   & 6     & CPITRNSL & CPI : Transportation & M\_110157302 & \multicolumn{1}{l}{CPI-U: transp} & \\
    113   & 6     & CPIMEDSL & CPI : Medical Care & M\_110157304 & \multicolumn{1}{l}{CPI-U: medical}&  \\
    114   & 6     & CUSR0000SAC & CPI : Commodities & M\_110157314 & \multicolumn{1}{l}{CPI-U: comm.}&  \\
    115   & 6     & CUSR0000SAD & CPI : Durables & M\_110157315 & \multicolumn{1}{l}{CPI-U: dbles}&  \\
    116   & 6     & CUSR0000SAS & CPI : Services & M\_110157325 & \multicolumn{1}{l}{CPI-U: services} & \\
    117   & 6     & CPIULFSL & CPI : All Items Less Food & M\_110157328 & \multicolumn{1}{l}{CPI-U: ex food} & \\
    118   & 6     & CUSR0000SA0L2 & CPI : All items less shelter & M\_110157329 & \multicolumn{1}{l}{CPI-U: ex shelter} & \\
    119   & 6     & CUSR0000SA0L5 & CPI : All items less medical care & M\_110157330 & \multicolumn{1}{l}{CPI-U: ex med} & \\
    120   & 5     & PCEPI & Personal Cons. Expend.: Chain Index & gmdc  & \multicolumn{1}{l}{PCE defl} & \\
    121   & 6     & DDURRG3M086SBEA & Personal Cons. Exp: Durable goods & gmdcd & \multicolumn{1}{l}{PCE defl: dlbes}&  \\
    122   & 6     & DNDGRG3M086SBEA & Personal Cons. Exp: Nondurable goods & gmdcn & \multicolumn{1}{l}{PCE defl: nondble} & \\
    123   & 6     & DSERRG3M086SBEA & Personal Cons. Exp: Services & gmdcs & \multicolumn{1}{l}{PCE defl: service} & \\
    \multicolumn{6}{l}{\textit{Group 8: Stock market}}  & \\
    124   & 5     & S\&P 500 & S\&P’s Common Stock Price Index: Composite & M\_110155044 & \multicolumn{1}{l}{S\&P 500}&  \\
    125   & 5     & S\&P: indust & S\&P’s Common Stock Price Index: Industrials & M\_110155047 & \multicolumn{1}{l}{S\&P: indust} & \\
    126   & 2     & S\&P div yield & S\&P’s Composite Common Stock: Dividend Yield &       & \multicolumn{1}{l}{S\&P div yield} & 0.14 \\
    127   & 5     & S\&P PE ratio & S\&P’s Composite Common Stock: Price-Earnings Ratio &       & \multicolumn{1}{l}{S\&P PE ratio} & \\
    128   & 1     & VXOCLSx & VXO   &       &  & 5.75\\
\label{table:data}
\end{longtable}
\end{ThreePartTable}
\endgroup
\end{landscape}

\section{On recurrent neural networks} \label{app:RNN}

The internal memory of a plain recurrent neural net (RNN) can be described as 
\begin{equation}
    \begin{aligned}
        & f_{t|L} = \Gamma (w' z_t + u' f_{t-1|L} + b)\\ 
        & f_{0|L} = 0,
    \end{aligned}
\label{eq:RNN}
\end{equation}
with parameters $(w,u,b)'$, where $w \in \mathbb{R}^{N \times p}$, $u \in \mathbb{R}^{p \times p}$ and $b \in \mathbb{R}^{p \times 1}$.\footnote{In this example, and throughout the paper, I assume an RNN with an architecture \emph{many-to-one}, where at each time step the network receives the inputs, updates the internal memory, and only delivers the output after all lags $l=1,...,L$ have been processed through the recursion equation. Other variations exist, such as the specification \emph{many-to-many}, in which the network delivers an output at each time step.} The final prediction of the RNN is $G(x_t; \Theta_h) = g \Big( f_{t|L} \Big)$, where $g$ is a linear function. Common candidates for $\Gamma$ are the sigmoid and tanh functions, and is applied element-wise.

The drawback of traditional RNNs is that they usually suffer from vanishing gradients that compromises the estimation process through a slow rate of improvement. Suppose that we seek to estimate an RNN with number of lags $L$. In brief terms, the estimation of the RNN involves computing the gradient of the loss function with respect to the parameters, which implies evaluating the gradient at every time step within sequences of observations of length $L$. If the parameters are significantly small (usually they are close to zero), the higher $L$, the smaller the contribution of observations sufficiently back in time, given the multiplicative effect of the chain rule and the fact that the derivative of the activation function is bounded by 1 (supposing the commonly used tanh or sigmoid functions). In other words, the model will not properly estimate long-term dependencies because the estimation process is compromised for sufficiently long sequences.\footnote{In practice, the vanishing gradients problem may also occur in significantly deep feed-forward networks. The use of the $ReLu$ activation function in these cases helps preventing the problem because its derivative is either $0$ or $1$.}

The RNN with long-short term memory units, i.e. the LSTM model, solves this problem by avoiding the gradients to be too small, which is key to explaining the long-memory feature usually attributed to LSTMs. The intuition behind this process relies on the existence of a cell state that turns out to be more stable than its counterpart in the traditional RNN, stabilizing the gradients as a consequence. This stability comes from the additive nature of the cell state, as well as the presence of filters that control the flow of information. These features together ensure suitable values for the gradient. For instance, if information from time step $l$ shouldn't be forgotten to predict $y_{t+h}$, the parameters of specific filters are estimated accordingly so that the gradient at the $l$-time step is sufficiently large to account for this information when updating the model parameters. Intuitively, this mechanism allows the information to effectively flow across time periods. For a more thorough description of RNNs, see \cite{Goodfellow-et-al-2016}.

\section{On the estimation of neural networks}
\label{app:estimation}

The parameters of neural network models are estimated by minimizing a loss function between the fitted and actual values over the in-sample period. In this application I use the mean squared error loss as specified in equation (\ref{eq:loss}). Given that the function $G$ is nonlinear with respect to the covariates $x_t$, the problem of minimizing the loss usually translates into optimizing a non-convex function. In these cases, iterative algorithms are more suitable than the classic optimization procedures applied to convex functions because of their properties that enforce the algorithm to rapidly converge to optima (\citealp{jain}). The literature on neural networks usually applies gradient descent as an optimization method. Gradient descent is based on the property that, to minimize a given function $\mathcal{L}$, one needs to move in the direction of the negative gradient, -$\Delta_{\Theta} \mathcal{L}(\Theta)$. The parameters are then updated iteratively, such that $\Theta_i = \Theta_{i-1} - \alpha \Delta_{\Theta} \mathcal{L}(\Theta_{i-1})$, where $\alpha$ is the learning rate, determining the size of the step, and $i$ is the iteration. In this study, I apply an extension of the gradient descent, called Adam optimization algorithm (\citealp{adam}), that features an adaptive learning rate.

The computation of the gradient may involve a single, random picked observation (stochastic gradient descent), a sub-group of observations (minibatch), or even all available observations (batch gradient descent). For the purpose of this analysis, the number of observations to be included, called batch size, is defined by grid search. Another important concept in machine learning is an epoch, defined as the number of passes of all observations through the algorithm, and not to be confounded with the number of iterations. At each time the gradient is computed, the algorithm updates the parameters, what defines an iteration. Hence, for the case of batch gradient descent the number of epochs coincides with the number of iterations. However for both stochastic gradient descent and minibatch methods, the number of iterations exceeds the number of epochs. The ultimate number of epochs is a hyperparameter selected by grid search.

The estimation process of neural networks, based on incremental updates of the parameters, means that the choice of the initial parameters is an important one. First, assigning equal weights to different nodes implies that they account for the same information and are therefore redundant. Random initialization is popular because it breaks the symmetry in the network. Second, one should avoid imposing too high or too low initial weight values in order to prevent the vanishing (or exploding) gradient problem, mentioned in Section \ref{app:RNN}. Modern approaches to parameter initialization rely on the idea that the variance of the activations (output of nodes) should be similar across layers (\citealp{hanin}). This literature suggests that parameters should therefore be randomly drawn from some zero-centered distribution with a specified variance, while biases are usually initialized with zeros. Common approaches are the Xavier, Glorot and He initializations (\citealp{glorot}, \citealp{he}). This application considers the Glorot initialization, in which initial weights are drawn from a specific uniform distribution.\footnote{Glorot initialization: $W \sim U \left [ - \frac{\sqrt{6}}{\sqrt{n_q + n_{q+1}}} ,  \frac{\sqrt{6}}{\sqrt{n_q + n_{q+1}}} \right ]$, where $n_q$ is the number of input units to layer $q$.}

The non-convexity usually encountered in neural networks tends to increase the sensitivity of the learning algorithm to initial values. This means that in practice the model delivers a slightly different prediction every time it is re-estimated, given the random initialization. A common solution adopted by the empirical literature is to repeat the estimation a (large) number of times and average out the predictions, which significantly reduces the variance of the prediction and consequently the uncertainty around initial values. Section \ref{sec:iv} studies how forecasts compare across models with different initializations and with the ensemble prediction in the context of the present empirical exercise.

\section{Model specification} \label{app:model_spec}

The cross-validation process is split into two stages. Stage 1 focuses on hyperparameters specific to the model's architecture, while stage 2 selects the hyperparameters related to the optimization procedure. For example, the number of nodes in the network would be selected in stage 1, while the batch size in stage 2. This method shrinks significantly the computational time compared to the option of selecting all hyperparameters at once. Moreover, previous tests (not reported) indicated that the relative performance of the models are not very sensitive to changes in the number of epochs or batch size. The grid search around these hyperparameters is nonetheless performed for robustness purposes in stage 2. I henceforth refer to ``specification'' as a particular selection of a set of hyperparameters.

During stage 1, I follow a step-by-step procedure: the FF-cpi and FF-pool models are estimated first, followed by the LSTM-pool and LSTM-all models. As explained below, the cross-validation on the FF-LSTM model only occurs in stage 2. Both FF-cpi and FF-pool are estimated over 64 different specifications, where I let vary the number of lags, the number of nodes in the feed-forward layer(s) and the number of hidden layers in the network, as indicated in Table \ref{table:hyper}.

Second, I use the optimal selection of number of nodes as estimated from the FF models as \emph{fixed} hyperparameters in the cross-validation of the LSTM models (recall that this model also includes a FF section stacked to the LSTM unit). Cross-validation is then performed over 32 different specifications for the LSTM models, including the number of lags, the number of hidden layers and the number of factors. Finally, I set as fixed in the FF-LSTM model the optimal specifications selected from the previous steps. More specifically, I set the number of lags $L$ in the feed-forward part of the model as equal to the optimal value from the FF-cpi, and the number of lags $L$ to be included in the LSTM unit as equal to the optimal value from the LSTM-pool model, as well as the optimal values of nodes and hidden states. The reason behind these choices relies on the similarities between the underlying model structures. The strategy of fixing hyperparameters based on optimal values of nested models facilitates the comparison between models and significantly decreases the computational time. Finally, during stage 2, the hyperparameters related to the optimization process are allowed to vary for all models.

Each model specification is evaluated over a so called validation sample. First consider splitting the full sample with $T$ observations between an in-sample period, of size $R$, and an out-of-sample period, of size $P$, such that $T = R + P + h -1$. The in-sample period is further split into two consecutive sets, the training and validation samples. Each specification is then estimated over the training sample and evaluated over the validation sample.\footnote{The forecast performance is measured as the root mean squared error.} The corresponding performance is used to differentiate between specifications.

\begin{figure}[h!]
    \centering
    \caption{Illustrative setup of the cross-validation and out-of-sample forecasting}
    \includegraphics[width=15cm]{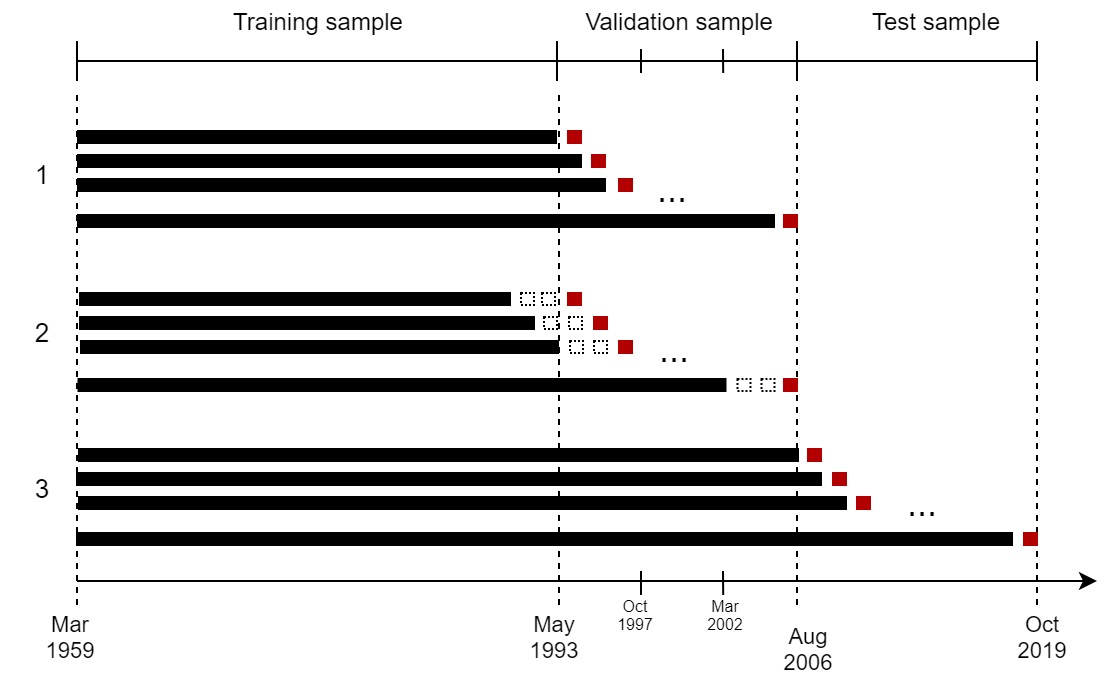}
    \caption*{\small Part 1 (top) depicts the cross-validation over the validation sample for the one-step ahead model. For each specification of hyperparameters, the model is trained over the training sample, and at each step over the validation sample one more data point is added for estimation. The performance over the validation sample is then given by the average performance over three sub-samples: May-1993 to October-1997, May-1993 to March-2002, May-1993 to July-2006. Part 2 (middle) depicts the same cross-validation procedure but for a three-steps ahead model. Note that the predictions cover all data points in the validation set. Finally, part 3 (bottom) illustrates the out-of-sample forecasting for a one-step ahead model, computed for the optimal model selected via cross-validation. The models are re-estimated every 48 months for both the cross-validation and out-of-sample performance.}
    \label{fig:cross_val}
\end{figure}

More specifically, the cross-validation exercise is implemented as follows: (i) split the data into consecutive samples: training, validation (three different lengths) and out-of-sample set, known as test sample in the machine learning jargon;\footnote{Choosing the length of each sample can be quite arbitrary and ultimately depends on each application. Here I split the data such that approximately $60\%$ of the total sample is devoted to training only, $20\%$ to validation and $20\%$ to testing. The precise splits are shown in Figure \ref{fig:cross_val}.} (ii) for each specification, estimate the model over the training sample and predict over the validation sample; (iii) repeat this process $140$ times, and compute the average prediction, defined as the series that averages out the predictions of the $140$ series at each point in time; (iv) measure the performance of the average prediction over each of the three sections of the validation sample; and (v) choose the specification with best average performance over the three sections of the validation sample. The out-of-sample performance is then obtained by evaluating the best specification on unseen data, in which case the final prediction is the average over $1400$ different prediction series.\footnote{The number of iterations for cross-validation is much smaller than the one used for out-of-sample performance as a way of minimizing the computational time given the high number of specifications to estimate. The choice is nonetheless somewhat arbitrary, and the number of repetitions were ultimately set such that it is a multiple of the number of available processors (28).}

I use a modified version of the more traditional k-fold cross-validation to account for time series characteristics. More specifically, at each time step during cross-validation and out-of-sample performance the estimation set expands by one observation and the prediction is compared with the actual value. By estimating and cross-validating the model on consecutive samples it is possible to avoid the look-ahead bias, since the performance is measured only on future data. In addition, I evaluate each specification over three nested sub-samples of the validation sample, the longest being the full validation set, and choose the model with best average performance over the splits. This is a simple way to add robustness to the analysis, since it minimizes the chances of sample-dependent results. In practice, the choice of the number of splits is quite arbitrary. Here the splits are selected such that the minimum sub-sample size comprises approximately four years of observations.

\clearpage
\begin{landscape}

\begin{table}[hp!] \footnotesize
\centering 
    \begin{threeparttable}
    \caption{Candidate and optimal values for hyperparameters}
    \begin{tabular}{lcccccccccccccc}
    \toprule
          & \multicolumn{2}{c}{FF-cpi} &       & \multicolumn{2}{c}{FF-pool} &       & \multicolumn{2}{c}{LSTM-pool} &       & \multicolumn{2}{c}{LSTM-all} &       & \multicolumn{2}{c}{FF-LSTM} \\
    \midrule
          & Candidates & Optimal &       & Candidates & Optimal &       & Candidates & Optimal &       & Candidates & Optimal &       & Candidates & Optimal \\
\cmidrule{2-3}\cmidrule{5-6}\cmidrule{8-9}\cmidrule{11-12}\cmidrule{14-15}    \textit{Stage 1} &       &       &       &       &       &       &       &       &       &       &       &       &       &  \\
    \quad lags $L$  & 6,12,24,48 & 24    &       & 6,12,24,48   & 48   &       & 6,12,24,48   & 48   &       & 6,12,24,48 & 48    &       &    & 24, 48 \\
    \quad nodes $n$  & 16, 32, 64, 128 & 128   &       & 16, 32, 64, 128 & 128   &       & 16, 32, 64, 128   & 128   &       & 16, 32, 64, 128   & 128   &       &   & 128 \\
    \quad layers $Q$ & 1,2,3,4 & 4     &       & 1,2,3,4 & 3     &       & 3, 4  & 4     &       & 3, 4  & 4     &       &   & 4 \\
    \quad $f_{t|L}$-size $p$ &   &   &       &    &   &       & 2,4,6,8 & 2     &       & 2,4,6,8 & 2     &       &    & 2 \\
    \# parameters &     & 80,513 &       &    & 758,273 &       &     & 51,017 &       &      & 51,097 &       &     & 81,737 \\
    \hline
    \textit{Stage 2} &       &       &       &       &       &       &       &       &       &       &       &       &       &  \\
    \quad epochs & 200,400,600 & 200   &       & 200,400,600 & 400   &       & 200,400,600 & 400   &       & 200,400,600 & 400   &       & 200,400,600 & 400 \\
    \quad batch & 128, \textit{max} & 128   &       & 128, \textit{max} & 128   &       & 128, \textit{max} & \textit{max} &       & 128, \textit{max} & \textit{max} &       & 128, \textit{max} & 128 \\
    \bottomrule
    \end{tabular}%
    \begin{tablenotes}
    \small
    \item The table reports the candidate values for each hyperparameter as well as the optimal value selected by cross-validation. The total number of parameters is specific to the CPI target (it is similar for the case of PCE although not exactly the same). The optimal hyperparameters of the FF-cpi and LSTM-pool selected in stage 1 are applied to the FF-LSTM model. Lag $L$ implies that all lags up to $L$ are included in the model. The batch size specified as \emph{max} corresponds to the batch gradient descent method. Not applicable cases are left blank. \vspace{5mm}
    \end{tablenotes}
    \label{table:hyper}%
    \end{threeparttable}
\end{table}%

\end{landscape}

\clearpage

\begin{table}[hp!] \small
\centering 
    \begin{threeparttable}
    \caption{Hyperparameter values - Small networks}
    \begin{tabular}{lccccc}
    \toprule
          & \multicolumn{1}{c}{FF-cpi} & \multicolumn{1}{c}{FF-pool} & \multicolumn{1}{c}{LSTM} & \multicolumn{1}{c}{LSTM-all} & \multicolumn{1}{c}{FF-LSTM} \\
    \textit{CPI} &       &       &       &       &  \\
    \midrule
    \quad inputs $M$ & \multicolumn{1}{c}{10} &  &  &  & \multicolumn{1}{c}{10} \\
    \quad inputs $N$ &  & \multicolumn{1}{c}{118} & \multicolumn{1}{c}{118} & \multicolumn{1}{c}{128} & \multicolumn{1}{c}{118} \\
    \quad lags $L$ & \multicolumn{1}{c}{24} & \multicolumn{1}{c}{48} & \multicolumn{1}{c}{48} & \multicolumn{1}{c}{48} & \multicolumn{1}{c}{24, 48} \\
    \quad nodes $n$ & \multicolumn{1}{c}{4} & \multicolumn{1}{c}{4} & \multicolumn{1}{c}{4} & \multicolumn{1}{c}{4} & \multicolumn{1}{c}{4} \\
    \quad layers $Q$ & \multicolumn{1}{c}{4} & \multicolumn{1}{c}{4} & \multicolumn{1}{c}{4} & \multicolumn{1}{c}{4} & \multicolumn{1}{c}{4} \\
    \quad factors $p$ &  &  & \multicolumn{1}{c}{2} & \multicolumn{1}{c}{2} & \multicolumn{1}{c}{2} \\
    \# parameters & \multicolumn{1}{c}{      1,029 } & \multicolumn{1}{c}{      22,725 } & \multicolumn{1}{c}{      1,045 } & \multicolumn{1}{c}{      1,125 } & \multicolumn{1}{c}{      2,005 } \\
    \textit{PCE} &       &       &       &       &  \\
    \midrule
    \quad inputs $M$ & \multicolumn{1}{c}{3} &  &  &  & \multicolumn{1}{c}{3} \\
    \quad inputs $N$ &  & \multicolumn{1}{c}{125} & \multicolumn{1}{c}{125} & \multicolumn{1}{c}{128} & \multicolumn{1}{c}{125} \\
    \quad lags $L$ & \multicolumn{1}{c}{24} & \multicolumn{1}{c}{48} & \multicolumn{1}{c}{48} & \multicolumn{1}{c}{48} & \multicolumn{1}{c}{24, 48} \\
    \quad nodes $n$ & \multicolumn{1}{c}{4} & \multicolumn{1}{c}{4} & \multicolumn{1}{c}{4} & \multicolumn{1}{c}{4} & \multicolumn{1}{c}{4} \\
    \quad layers $Q$ & \multicolumn{1}{c}{4} & \multicolumn{1}{c}{4} & \multicolumn{1}{c}{4} & \multicolumn{1}{c}{4} & \multicolumn{1}{c}{4} \\
    \quad factors $p$ &  &  & \multicolumn{1}{c}{2} & \multicolumn{1}{c}{2} & \multicolumn{1}{c}{2} \\
    \# parameters &          357  &       24,069  &       1,101  &       1,125  &       1,389  \\
    \bottomrule
    \end{tabular}%
    \begin{tablenotes}
    \small
    \item The table reports the hyperparameters of a set of ``smaller'' models, where the number of nodes in hidden layers is fixed at $4$ and not cross-validated. The other hyperparameters are maintained the same as in table \ref{table:hyper}. Not applicable cases are left blank. \vspace{5mm}
    \end{tablenotes}
    \label{table:hyper_small}%
    \end{threeparttable}
\end{table}%

\clearpage

\section{Extended results for Section \ref{sec:OFS}} \label{app:OFS_app}

\begin{figure}[h!]
\caption{Local forecast accuracy and Predictions for CPI inflation - Horizons 3 and 6}
\centering
\begin{adjustbox}{max width=5\linewidth,center}
\subfloat[\small Out-of-sample 1993-2006  \vspace{-2mm} ]{{\includegraphics[width=1.05\textwidth]{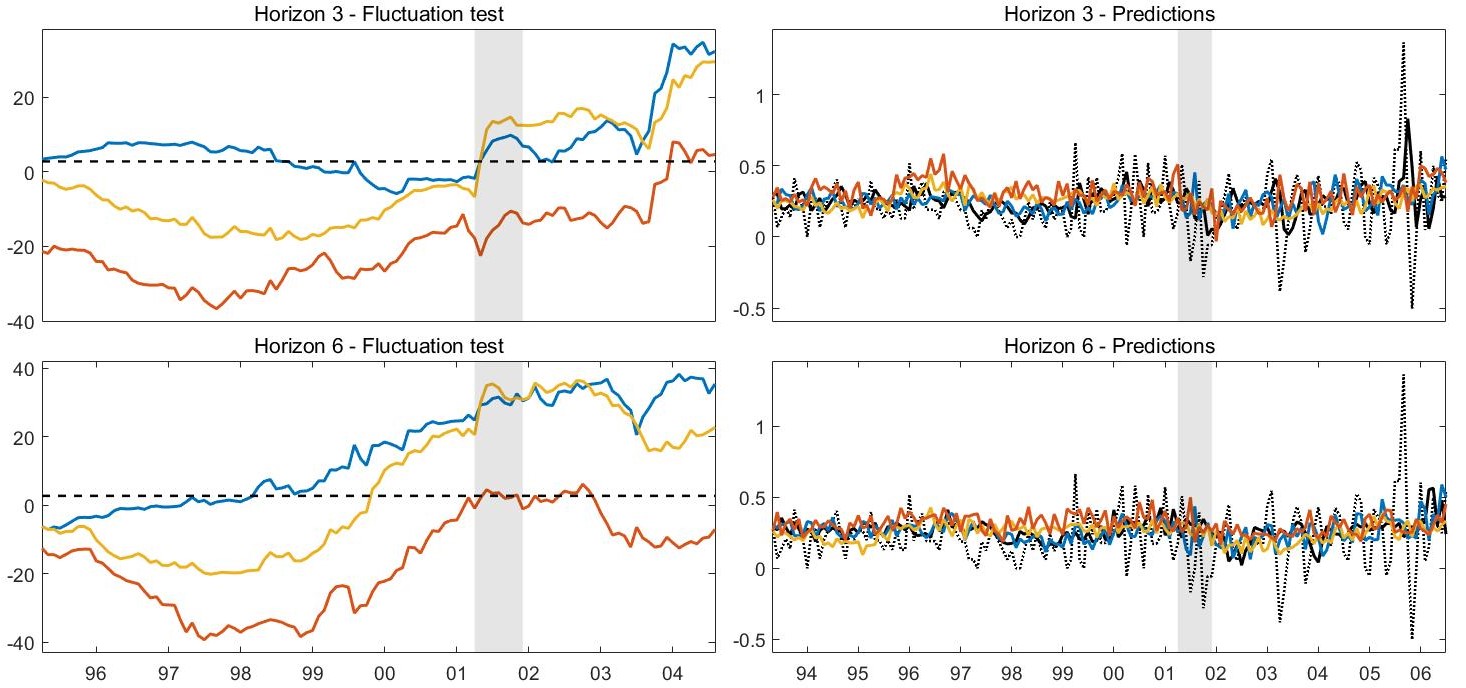}}}
\end{adjustbox}
\begin{adjustbox}{max width=5\linewidth,center}
\subfloat[\small Out-of-sample 2006-2019 \vspace{-2mm}]{{\includegraphics[width=1.05\textwidth]{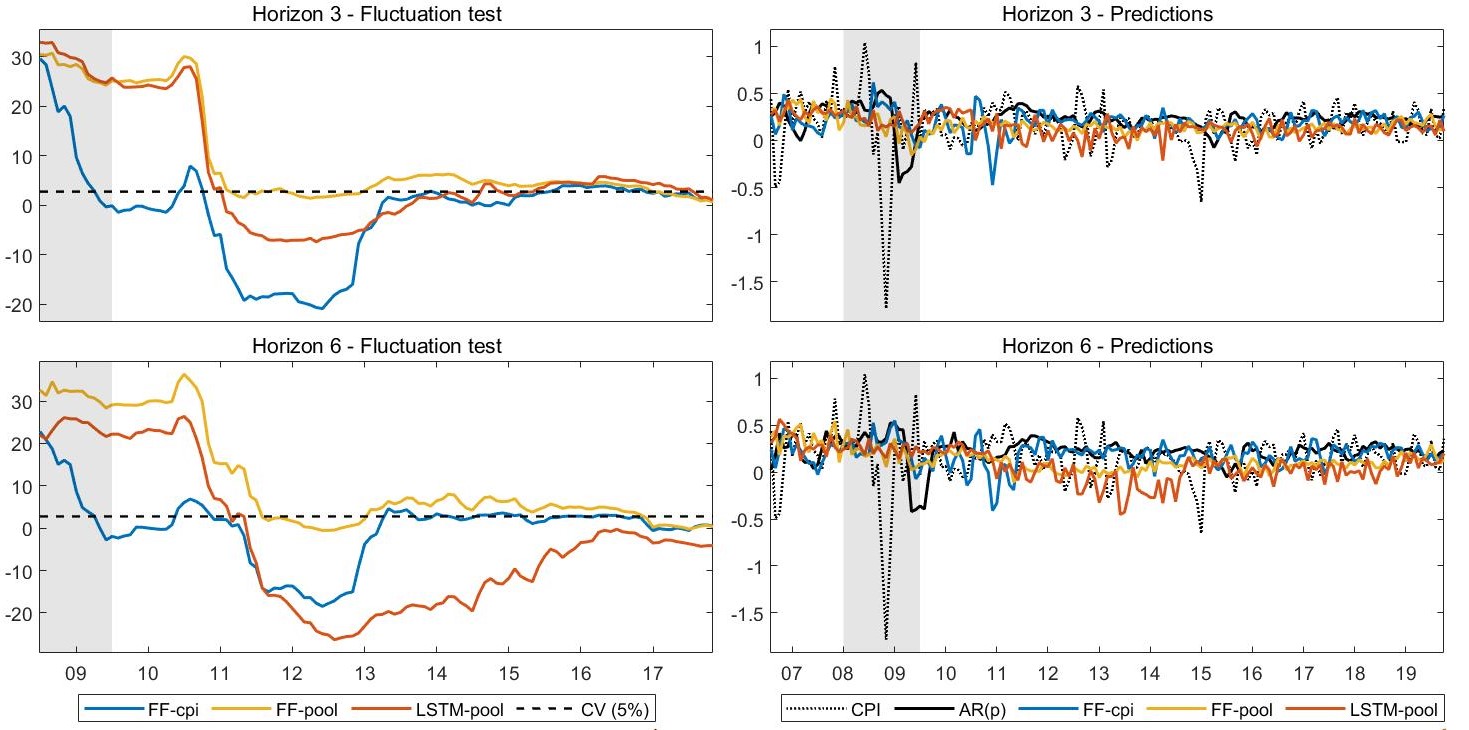}}}%
\end{adjustbox}
\caption*{\small First column: Test statistics according to the Fluctuation Test from \cite{rossi}; values above the dashed line indicate that the underlying model presents better forecast accuracy than the $AR(p)$ benchmark over a window of fixed size at the $5\%$ confidence level. Second column: Predictions of selected models; the scale is $100 \times$ log differences. Gray areas are NBER recessions.}
\label{fig:ft_app}
\end{figure}

\begin{table}[h!] \centering \footnotesize
\begin{threeparttable}
  \caption{Out-of sample forecast performance for PCE inflation}
   \begin{tabular}{lccccccccccc}
   \toprule
          & \multicolumn{5}{c}{1993-2006}         &       & \multicolumn{5}{c}{2006-2019} \\
\cmidrule{2-6}\cmidrule{8-12}          & 1     & 3     & 6     & 12    & 24    &       & 1     & 3     & 6     & 12    & 24 \\
    \multicolumn{6}{l}{\textit{Benchmark models}} &       &       &       &       &       &  \\
    \midrule
    \quad UCSV  & \textbf{1.05} & \textbf{1.14} & \textbf{1.04} & \textbf{1.04} & \textbf{0.89**} &       & 1.06  & 0.99  & 1.02  & 1.03  & \textbf{1.05} \\
    \quad Fact  & \textbf{1.01} & \textbf{1.00} & \textbf{1.01} & 1.10  & \textbf{1.04} &       & \textbf{0.98} & \textbf{1.02} & \textbf{1.02} & \textbf{1.03} & \textbf{1.04} \\
    \quad RF    & \textbf{0.94*} & \textbf{1.04} & \textbf{1.03} & \textbf{0.91***} & \textbf{0.89**} &       & \textbf{1.01} & \textbf{0.87***} & \textbf{0.88***} & \textbf{0.93***} & \textbf{0.90***} \\
    \quad LASSO  & \textbf{0.97*} & \textbf{1.04} & \textbf{1.06} & \textbf{1.03} & \textbf{1.01} &       & \textbf{0.93***} & \textbf{0.95***} & \textbf{0.94***} & \textbf{0.93***} & \textbf{0.92**} \\
    \quad Ridge & \textbf{1.00} & \textbf{1.10} & 1.14  & \textbf{1.00} & \textbf{0.91***} &       & \textbf{1.04} & 1.18  & 1.30  & 1.17  & 1.33 \\
    \quad ENet & \textbf{0.96**} & \textbf{1.03} & \textbf{1.07} & 1.03  & \textbf{1.00} &       & \textbf{0.94***} & \textbf{0.92***} & \textbf{0.94***} & \textbf{0.91***} & \textbf{0.90***} \vspace{2mm} \\
    \multicolumn{6}{l}{\textit{Neural network models}} &       &       &       &       &       &  \\
    \midrule
    \multicolumn{6}{l}{\textit{Large architecture}} &       &       &       &       &       &  \\
    \quad FF-cpi & \textbf{0.91**} & \textbf{1.02} & \textbf{1.01} & \textbf{1.03} & 1.22  &       & \textbf{1.01} & \textbf{0.93***} & \textbf{0.92**} & \textbf{0.94**} & \textbf{0.95**} \\
    \quad FF-pool & \textbf{1.04} & \textbf{1.12} & \textbf{1.10} & \textbf{0.96} & \textbf{0.98} &       & \textbf{1.01} & \textbf{0.84***} & \textbf{0.83***} & \textbf{0.90**} & \textbf{0.99} \\
    \quad LSTM-pool & \textbf{0.93*} & \textbf{1.11} & 1.21  & \textbf{0.94*} & \textbf{0.99} &       & \textbf{1.00} & \textbf{0.89**} & \textbf{0.98} & \textbf{0.91**} & \textbf{0.92**} \\
    \quad LSTM-all & \textbf{0.93*} & \textbf{1.09} & 1.17  & \textbf{0.91**} & \textbf{0.97} &       & \textbf{0.99} & \textbf{0.89**} & \textbf{0.98} & \textbf{0.91***} & \textbf{0.92**} \\
    \quad FF-LSTM & \textbf{0.90**} & \textbf{1.01} & \textbf{1.01} & \textbf{1.00} & 1.11  &       & \textbf{1.00} & \textbf{0.90***} & \textbf{0.90***} & \textbf{0.91***} & \textbf{0.90***} \\
    \multicolumn{6}{l}{\textit{Small architecture}} &       &       &       &       &       &  \\
    \quad FF-cpi & \textbf{0.96} & \textbf{1.04} & \textbf{1.03} & \textbf{1.01} & \textbf{1.05} &       & \textbf{1.03} & \textbf{0.90***} & \textbf{0.90***} & \textbf{0.92***} & \textbf{0.95**} \\
    \quad FF-pool & \textbf{1.05} & \textbf{1.14} & \textbf{1.11} & \textbf{0.96*} & \textbf{0.90***} &       & \textbf{1.06} & \textbf{0.86***} & \textbf{0.87***} & \textbf{0.97} & \textbf{1.04} \\
    \quad LSTM-pool & \textbf{0.98} & \textbf{1.14} & 1.25  & \textbf{0.98} & \textbf{0.99} &       & \textbf{1.02} & \textbf{0.86***} & \textbf{0.88**} & \textbf{0.90***} & \textbf{0.92***} \\
    \quad LSTM-all & \textbf{0.98} & \textbf{1.13} & 1.20  & \textbf{0.96} & \textbf{0.96} &       & \textbf{1.02} & \textbf{0.87***} & \textbf{0.88**} & \textbf{0.90***} & \textbf{0.92***} \\
    \quad FF-LSTM & \textbf{0.92**} & \textbf{1.04} & \textbf{1.04} & \textbf{0.93**} & \textbf{0.89***} &       & \textbf{0.98} & \textbf{0.84***} & \textbf{0.85***} & \textbf{0.86***} & \textbf{0.89***} \\
    \bottomrule
    \end{tabular}%
    \begin{tablenotes}
    \small
    \item The table presents the loss ratios with respect to the $AR(p)$ model for horizons $h=1,3,6,12,24$ and two OOS periods for models forecasting PCE inflation. The loss function is the RMSE. *, **, *** denote significance of the one-sided DM test at a $10\%$, $5\%$ and $1\%$ levels respectively. Models retained in the $75\%$ MCS are in bold.
    \vspace{2mm}
    \end{tablenotes}
    \label{table:rmse_pce}
\end{threeparttable}
\end{table}%

\clearpage

\section{Extended results for Section \ref{sec:iv}} \label{app:iv_app}

\begin{figure}[h!]
    \centering
    \caption{Distribution of the CW statistic over different initializations}
    \includegraphics[width=16cm]{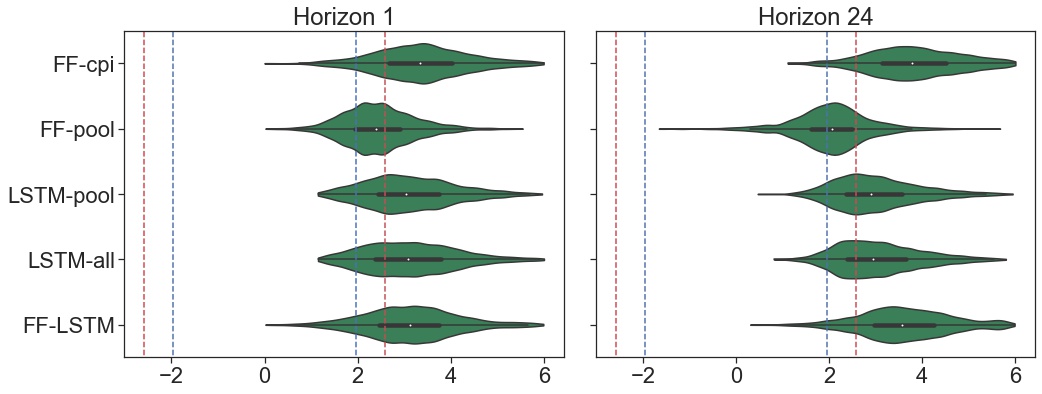}
    \caption*{\small \cite{clark_west07} (CW) test statistic $\Delta_{k,h}$, for $k=1,...,1400$ and horizons $h=1,24$. Positive values imply better forecast performance of the ensemble prediction over predictions from a particular initialization. Vertical dashed lines are Gaussian critical values at the $5\%$ (blue) and $1\%$ (red) significance levels. Values are truncated at 6 for visualization purposes.}
\label{fig:DMadj}
\end{figure}

\begin{figure}[h!]
    \centering
    \caption{Distribution of the DM statistic over different initializations, Other horizons}
    \includegraphics[width=16cm]{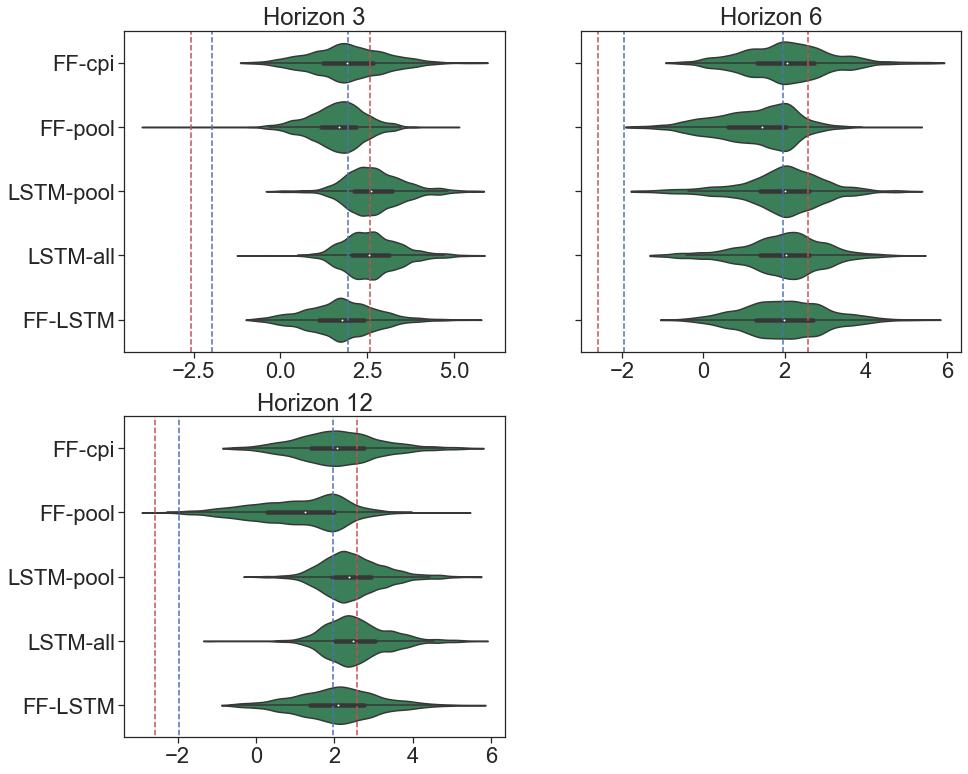}
    \caption*{\small DM test statistic $\Delta_{k,h}$, for $k=1,...,1400$ and horizons $h=3, 6, 12$. Positive values imply better forecast performance of the ensemble prediction over predictions from a particular initialization. Vertical dashed lines are Gaussian critical values at the $5\%$ (blue) and $1\%$ (red) significance levels. Values are truncated at 6 for visualization purposes.}
\label{fig:DM-other-hor}
\end{figure}

\begin{figure}
    \centering
    \caption{Distributions of the loss-differential mean and standard deviation over different initializations, FF-cpi and FF-pool models}
    \includegraphics[width=16cm]{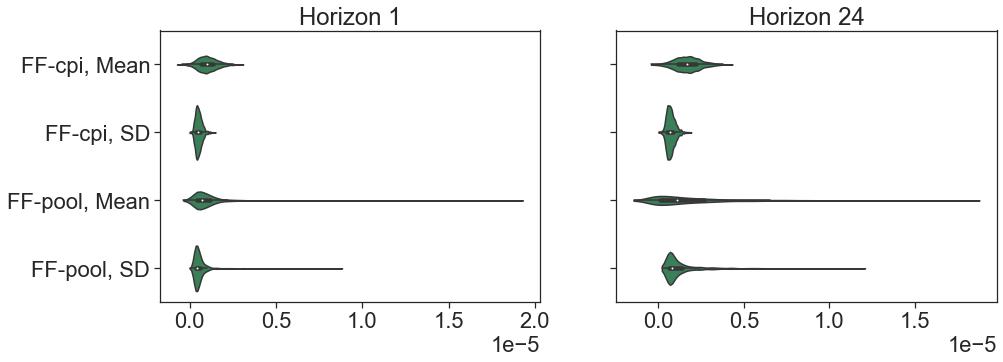}
    \label{fig:lossdiff-dist}
\end{figure}

\clearpage

\section{Benchmark specifications} \label{app:bench}

Consider the inflation series $\pi_t = log(P_t) - log(P_{t-1})$, where $P_t$ is the price index at time $t$.

\subsection{Autoregressive model $AR(p)$}

I estimate an autoregressive ($AR$) model of order $p$ for each horizon, where the $h$ step ahead forecast is given by
\begin{equation*}
    \widehat{\pi}_{t+h|t} = \widehat{c}_h + \widehat{\Phi}_{1,h} \pi_{t} + ... + \widehat{\Phi}_{p,h} \pi_{t-(p-1)}.
\end{equation*}
The model is estimated by least squares and $p$ is chosen by Bayesian Information Criterion (BIC), assuming maximum order of 4.

\subsection{Unobserved components with stochastic volatility (UCSV)}

A second benchmark is the UCSV model from \cite{stock07},
\begin{align*}
    \pi_t &= \tau_t + e^{h_t/2} \varepsilon_t \\
    \tau_t &= \tau_{t-1} + u_t \\
    h_t &= h_{t-1} + v_t
\end{align*}
where $\{\varepsilon_t\} \sim iid \mathcal{N}(0,1)$, $\{u_t\} \sim iid \mathcal{N}(0,\omega^2_{\tau})$, and $\{v_t\} \sim iid \mathcal{N}(0,\omega^2_{h})$. The state processes are initialized with $\tau_1 \sim \mathcal{N}(0,V_{\tau})$ and $h_1 \sim \mathcal{N}(0,V_{h})$, where $V_{\tau} = V_h = 0.12$. It is assumed independent inverse-gamma priors for $\omega^2_{\tau}$ and $\omega^2_h$. The model is estimated using Markov Chain Monte Carlo (MCMC) methods, and the $h$-step ahead forecast is given by
\begin{equation*}
    \widehat{\pi}_{t+h|t} = \widehat{\tau}_t
\end{equation*}

\subsection{Factor-augmented Distributed Lag (FADL) model, the ``Factor model''}

I specify a FADL$(p)$ model for each horizon $h$, where the $h$ step ahead forecast is given by
\begin{equation*}
    \widehat{\pi}_{t+h|t} = \widehat{\beta}_{0,h} + \sum_{i=0}^{p-1} \widehat{\beta}_{i+1,h} \pi_{t-i} + \sum_{j=1}^{r} \sum_{i=0}^{p-1} \widehat{\gamma}_{j,h} f_{j,t-i}.
\end{equation*}
The factors are estimated by principle components applied to the standardized data $\mathcal{Z}$. The number of factors $r$ as well as the number of lags $p$ are jointly selected by BIC, with maximum $r$ and $p$ set to 4. The model is estimated by least squares.

\subsection{Random Forest (RF)}

Random forests build on the concept of tree-based regression and classification models, or CART (\citealp{breiman84}; \citealp{breiman2001}). Here I provide a brief overview, and refer to \cite{hastie01statisticallearning} for a general introduction of the method.

The idea of tree-based models is essentially to split the predictor space $x_t$ into non-overlapping regions and fit a constant in each one. If the correspondent optimization criterion to choose splits is to minimize the sum of squares residuals, then the constant is simply the average of the target variable in each region. The regions are selected in a greedy manner, in which two regions are first selected out of the original data, then two other regions are selected from each of the first two, and so on. If estimated in the context of random forests, the trees can grown deep, where the minimum number of observations in each region is usually set to a small number.\footnote{I followed default values from the \texttt{randomForest} package in \texttt{R}, and set the minimum number of observations per region to 5 and the number of trees in the random forest to 500.}

I consider the predictor set $x_t = (z_t,...,z_{t-(L-1)},w_t,...,w_{t-(L-1)})'$, where $L$ is set to 4. The $h$ step ahead forecast of a single tree with $M$ regions $R_m$ is
\begin{equation*}
    \widehat{\pi}_{t+h|t}^{\mathcal{T}} = \sum_{m=1}^M \widehat{c}_m \text{ } \mathbf{1} \left[x_t \in R_m; \text{ } \Theta_h^{\mathcal{T}} \right],
\end{equation*}
where $\mathbf{1}$ is the indicator function and $\Theta_h^\mathcal{T}$ are the parameters of the tree.

Due to their hierarchical nature, trees can be very noisy, as a slight change in predictor values may lead to very different estimates. Forests successfully decrease the variance of tree predictions by combining many (de-correlated) trees in an ensemble fashion. More precisely, the prediction of a forest is
\begin{equation*}
    \widehat{\pi}_{t+h|t} = \frac{1}{B} \sum_{b=1}^B \widehat{\pi}_{t+h|t}^{\mathcal{T}_b},
\end{equation*}
where $\mathcal{T}_b$ is a tree (i) estimated on a (block) bootstrap sample $b$ of the original data, and (ii) considering only a subset $x_{t,b} \in x_t$ of the original predictor set. 

\subsection{Shrinkage models - LASSO, Ridge and Elastic-net}

Shrinkage models yield a $h$ step ahead prediction of the form $\widehat{\pi}_{t+h|t} = \widehat{\beta}_h x_t$, where the coefficients minimize a penalized residual sum of squares,
\begin{equation*}
    \widehat{\beta}_h = \underset{\beta}{argmin} \text{ } \sum_{t=1}^{T-h} \left( \pi_{t+h} - \beta_h' x_t \right)^2 + p(\beta_h,\lambda).
\end{equation*}
The tuning parameter $\lambda \geq 0$ controls the amount of shrinkage and is selected by BIC. I consider the predictor set $x_t = (z_t,...,z_{t-(L-1)},w_t,...,w_{t-(L-1)})'$, where $L$ is set to 4, and three variations of the penalty function $p(\beta_h,\lambda)$:
\begin{itemize}
    \item[1.] LASSO (Least Absolute Shrinkage and Selection Operator), with $p(\beta_h,\lambda) = \lambda \sum_{j=1}^{NT} |\beta_{j,h}|$;
    \item[2.] Ridge regression, with $p(\beta_h,\lambda) = \lambda \sum_{j=1}^{NT} \beta_{j,h}^2$; and
    \item[3.] Elastic-net, with $p(\beta_h,\lambda) = \alpha \lambda \sum_{j=1}^{NT} |\beta_{j,h}| + (1-\alpha) \lambda \sum_{j=1}^{NT} \beta_{j,h}^2$. The parameter $\alpha \in [0,1]$ is selected by BIC.
\end{itemize}

\end{document}